\documentclass[journal=aem,manuscript=article]{achemso}

\usepackage[version=3]{mhchem} 

\usepackage{chemformula} 
\makeatletter
\newcounter{chemeqn}
\newenvironment{chemequations}{\let\c@equation\c@chemeqn}{}
\makeatother
\usepackage{siunitx}
\usepackage{gensymb}
\usepackage{hyperref}
\DeclareSIUnit\angstrom{\text {Å}}
\DeclareUnicodeCharacter{2212}{-}

\author{Prakriti Kayastha}
\affiliation[NU]
{Department of Mathematics, Physics and Electrical Engineering, Northumbria University, Newcastle-upon-Tyne, NE1 8QH, United Kingdom}
\author{Giulia Longo}
\affiliation[NU]
{Department of Mathematics, Physics and Electrical Engineering, Northumbria University, Newcastle-upon-Tyne, NE1 8QH, United Kingdom}
\author{Lucy D. Whalley}
\affiliation[NU]
{Department of Mathematics, Physics and Electrical Engineering, Northumbria University, Newcastle-upon-Tyne, NE1 8QH, United Kingdom}
\email{l.whalley@northumbria.ac.uk}

\title{A first-principles thermodynamic model for the Ba-−Zr-−S system
in equilibrium with sulfur vapour}

\keywords{perovskite, solid-state, simulation, first-principles, thermodynamics, sulfur}

\begin{document}

\begin{abstract}
The chalcogenide perovskite \ce{BaZrS_3} has strong visible light absorption and high chemical stability, is nontoxic, and is made from earth-abundant elements. As such, it is a promising candidate material for application in optoelectronic technologies. However the synthesis of \ce{BaZrS_3} thin-films for characterisation and device integration remains a challenge. 
Here we use density functional theory and lattice dynamics to calculate the vibrational properties of elemental, binary and ternary materials in the Ba-Zr-S system. 
This is used to build a thermodynamic model for the stability of \ce{BaZrS_3}, \ce{BaS_x}, and \ce{ZrS_x} in equilibrium with sulfur gas, across a range of temperatures and sulfur partial pressures. 
We highlight that reaction thermodynamics are highly sensitive to sulfur allotrope and the extent of allotrope mixing.
We use our model to predict the synthesis conditions in which \ce{BaZrS3} and the intermediate binary compounds can form. 
At a moderate temperature of \SI{500}{\celsius} we find that \ce{BaS3}, associated with fast reaction kinetics, is stable at pressures above \SI{3E5}{\pascal}. We also find \ce{BaZrS3} is stable against decomposition into sulfur-rich binaries up to at least \SI{1E7}{\pascal}.
Our work provides insights into the chemistry of this promising material and suggests the experimental conditions required for the successful synthesis of \ce{BaZrS3}.\\
\textbf{Keywords:} perovskite, solid-state, simulation, first-principles, thermodynamics, sulfur
\end{abstract}


\section{Introduction}
   
Chalcogenide perovskites are a class of perovskite materials that have recently gained significant attention as lead-free alternatives for photovoltaic (PV) applications.\cite{sopiha2022chalcogenide,tiwari2021chalcogenide,jaramillo2019praise} 
They are environmentally stable and exhibit a range of desirable properties for optoelectronic applications, including defect tolerance,\cite{wu2021defect} strong dielectric screening\cite{ravi2021colloidal}, and high charge carrier mobility.\cite{osei2021examining} BaZrS$_3$ is the most studied material in this class due to its stability and high absorption coefficient.\cite{sun2015chalcogenide,nishigaki2020extraordinary} 
A wide bandgap in the range of \SIrange{1.8}{2.0}{\electronvolt}\cite{ju2017perovskite} makes it a suitable material for integration as a top cell absorber in Si-perovskite tandem applications.\cite{comparotto2020chalcogenide} 
In addition, the bandgap can be reduced through mixing on the Ba site,\cite{sharma2021bandgap} Zr site\cite{sharma2021bandgap,wei2020ti,Meng2016Alloying} or chalcogenide site,\cite{sadeghi2022new,odeh2023tuning} leading to the possibility of a single junction \ce{BaZrS3} solar cell.

Although \ce{BaZrS3} exhibits promising physical and chemical properties, a scalable synthesis method which avoids high temperatures and/or long reaction times remains an open challenge. 
During solid state synthesis, there are kinetic mass transport limitations when annealing the stable binary reactants; highly crystalline films are deposited at temperatures in the range of \SIrange{800}{1100}{\celsius}.\cite{Hahn1957unter,comparotto2020chalcogenide} 
Whilst the high kinetic barriers in chalcogenide perovskites might contribute to their thermal and chemical stability, the resulting high synthesis temperatures preclude thin-film growth on a photovoltaic device stack. 
   
To overcome kinetic limitations several groups are exploring the use of a barium polysulfide liquid \ce{BaS_x}, $x>3$ as a reaction   intermediate.\cite{sopiha2022chalcogenide,freund2022fabrication,vincent2023liquid,yang2023low}
This has lowered the temperature required for perovskite synthesis through the formation of a liquid flux of sulfur-rich barium polysulfide at around \SI{550}{\celsius}.\cite{binaryalloy2016,janz1976raman}
This approach allows for \ce{BaZrS3} synthesis using heat treatment at moderate temperatures in a sulfur-containing atmosphere.

Despite this recent progress in \ce{BaZrS3} synthesis at moderate temperatures, the underlying mechanisms and experimental conditions required for perovskite formation are not well understood. 
Initial experimental results have suggested there is a window of sulfur partial pressure within which the formation of \ce{BaZrS3} is favoured.\cite{sopiha2022chalcogenide} 
However it is unclear whether this window is determined by kinetic factors limiting the rate of perovskite formation,\cite{yang2023low} or the thermodynamic instability of \ce{BaZrS3} with respect to sulfur-rich binary materials \ce{BaS3} and \ce{ZrS3}. 


A complicating factor is that in the gas phase sulfur forms a range of S$_n$ allotropes at relative ratios which are highly sensitive to temperature and pressure,\cite{Dobbie1919The}
 and that this mixing can have a significant impact on predicted formation energies in the experimental conditions used for perovskite synthesis.\cite{jackson2016universal}
However the commonly made assumption is that the sulfur gas consists of a single species (most often \ce{S2} or \ce{S8}), which is determined by the temperature of the sulfur source.
Furthermore, experimental characterisation and control of the sulfur allotrope(s) formed is highly challenging.

\begin{figure}
    \centering
    \includegraphics[width=11cm]{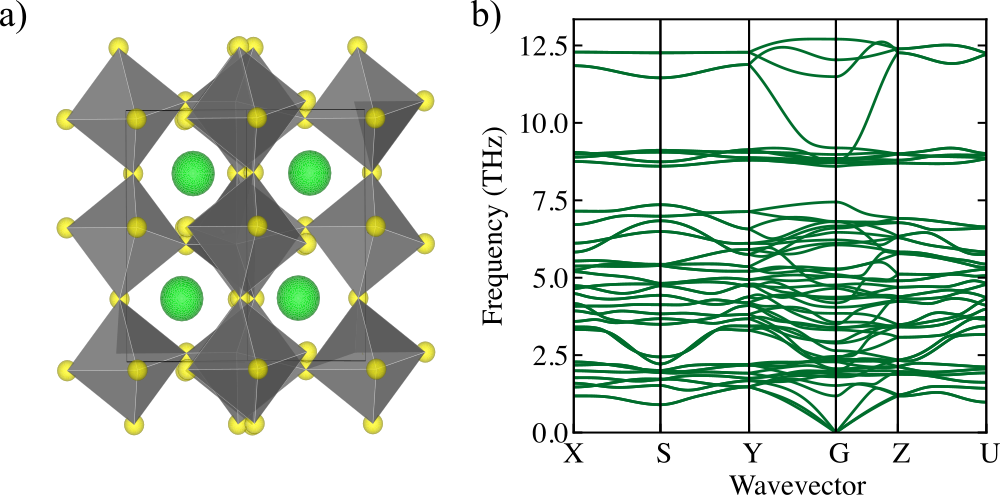}
    \caption{a) Relaxed crystal structures of \ce{BaZrS3} in the $Pnma$ orthorhombic perovskite phase. Ba atoms are green, S atoms are yellow and the Zr atoms, which lie in the centre of each octahedra (shaded grey), are not shown; b) Harmonic phonon bandstructure of \ce{BaZrS3} showing positive phonon modes across the Brillouin zone.}
    \label{fig:crystal_structures}
\end{figure}

In this work, we build a thermodynamic model for elemental, binary, and perovskite materials in the Ba-Zr-S system. 
We perform density functional theory and harmonic phonon calculations (Figure \ref{fig:crystal_structures}) to predict Gibbs free energies as a function of temperature and pressure.
Combining this with published experimental data\cite{chase1998nist} and a parameterised model\cite{jackson2016universal} for sulfur vapour allows us to incorporate the effects of sulfur partial pressure for the single-species \ce{S2} and \ce{S8} allotropes, in addition to an equilibrium sulfur gas with allotrope-mixing. 
Through this we demonstrate the impact of sulfur species on reaction thermodynamics, and predict the synthesis conditions required for \ce{BaZrS3} formation.
The first-principles data and material-agnostic post-processing code underlying this model are freely available online.\cite{ThermoPot,PaperRepo,NOMAD}

In our analysis we consider reaction pathways between the perovskite and competing binary and elemental phases. Ruddlesden-Popper (RP) phases are reported to form at high temperatures\cite{kayastha2023high,niu2019crystal,saeki1991preparation,hung1997ba3zr2s7} or for kinetically limited Zr precursors.\cite{yang2023low} RP phases are not examined here as we focus our attention on perovskite formation at moderate temperatures.

\section{Methodology}

\subsection{Classical thermodynamics}\label{sec:classical_thermodynamics}
We use classical thermodynamics to evaluate the stability of materials through the balance of free energies.
The feasibility of a spontaneous reaction at temperature $T$ and total pressure $P$ is determined by the change in Gibbs free energy $\Delta G$:
\begin{equation}
\Delta G_\mathrm{f}(T,P) = \sum_\mathrm{products}n_i\mu_i(T,p_i)- \sum_\mathrm{reactants}n_i\mu_i(T,p_i),
\end{equation}
where $n_i$ is the absolute change in stoichiometry, $\mu_i$ is the chemical potential and $p_i$ is the partial pressure of each product or reactant. 
For reactions with $\Delta G<0$, the reaction is energy releasing and thermodynamically favourable. 
We note that meeting this condition is not necessarily sufficient for forming a given product over a reasonable timescale; kinetic factors may inhibit the reaction. 

We assume that solids are incompressible and consider vibrational contributions to entropy only, giving the following expression for Gibbs free energy at finite temperature and pressure:\cite{Jackson2013Oxidation} 
\begin{equation}
    \mu_i(T,P) = E^\text{DFT} +E^\text{ZP} + \int_{0}^T C_p(T) \text{d}T + PV \\
    - TS_\mathrm{vib.}(T)
\end{equation}
Density functional theory (DFT) is used to calculate the total electronic energy $E^\text{DFT}$ and equilibrium volume $V$ for a static crystal at \SI{0}{\pascal}. We use first-principles lattice dynamics to calculate the zero point energy $E^\text{ZP}$, vibrational entropy $S_\mathrm{vib.}(T)$ and heat capacity $C_p(T)$. Note that $C_p=C_v$ for incompressible solids.

We assume that \ce{S2} and \ce{S8} follow the ideal gas law, giving:\cite{Jackson2013Oxidation}
\begin{equation}
    \mu_i(T,p_i) = E^\text{DFT} +E^\text{ZP} + [H^\theta - H^0] - \int_{T^\theta}^T C_p(T,p_i) \text{d}T \\+ RT \text{ln} [p_i/p_i^\theta] - TS(T,p_i^\theta).
\end{equation}
The NIST-JANAF thermochemical data tables provide experimental data for the standard enthalpy $[H^\theta - H^0]$, heat capacity $C_p(T,p_i)$ and entropy $S(T,p_i^\theta)$.\cite{chase1998nist} $\theta$ denotes the reference state used in the tables ($T=$\SI{298}{\kelvin}, $P=$\SI{1E5}{\pascal}). $E^\text{ZP}$ values are taken from the NIST Computational Chemistry Comparison and Benchmark Database.\cite{CCCBD}
We reference the $E_\text{DFT}$ value for \ce{S2} against $E_\text{DFT}$ for \ce{S8} using the energy difference employed to construct the mixed allotrope model in Reference \citenum{jackson2016universal}. This energy difference is calculated a hybrid PBE0 functional\cite{carlo1999toward} and is in close agreement with reference data.\cite{chase1998nist} Calculations with the SCAN,\cite{sun2015strongly} PBEsol\cite{perdew2008restoring} and HSE06\cite{krukau2006influence} functionals overestimate the energy difference leading to a coexistence curve that is significantly too high in temperature (see Figures S28 and S29).
To compare calculated $\mu$ values across sulfur allotropes, we report $\mu$ on a per-atom basis, i.e. $\mu_{S_8}/8$ for \ce{S8} and  $\mu_{S_2}/2$ for \ce{S2}.

To account for allotrope mixing, we applied a published first-principles model for the chemical potential of sulfur gas.\cite{jackson2016universal} This model considers 13 low-energy allotropes of sulfur (\ce{S2}--\ce{S8}) and is parameterised using first-principles lattice dynamics with the hybrid PBE0 functional.\cite{carlo1999toward}
We evaluated and visualised thermodynamic potentials and sulfur model using our in-house code \ce{ThermoPot}.\cite{ThermoPot} This analysis is available in an online repository.\cite{PaperRepo} \ce{ThermoPot} is a general-purpose and material-agnostic code which post-processes first-principles data.
We identified low energy materials in the Ba-Zr-S system using total energies reported on the Materials Project.\cite{jain2013commentary} 
More details on the database search are provided in the Supplementary Information.

\subsection{Quantum chemical calculations}

First-principles calculations were carried out with the all-electron numerical atom-centered orbital code FHI-aims. All calculations were performed using the default `tight' basis set in FHI-aims, which extends the minimal set of occupied orbitals with 6 additional functions.\cite{blum2009ab} For reciprocal space sampling a Monkhorst-Pack grid was used with a minimum $k$-spacing of \SI{0.5}{\per\angstrom}.

Equilibrium crystal structures were obtained using a parametrically constrained geometry relaxation implemented in the AFLOW Library interface to FHI-aims.\cite{lenz2019parametrically} Geometry relaxations for solid materials were performed using the generalized gradient approximated (GGA) PBEsol\cite{perdew2008restoring} functional. The structures were relaxed until the maximum force component was below \SI{5e-3} {\eV\per\angstrom}. We do not include a correction for van der Waals interactions, but note that our calculated value for interlayer spacing in \ce{ZrS2} (\SI{3.63}{\angstrom}) is within 0.5\% of the experimentally reported value at room temperature (\SI{3.65}{\angstrom}, ICSD entry 76037). The initial geometries for the sulfur gas species (\ce{S2} and \ce{S8}) were obtained from the NIST Computational Chemistry Comparison and Benchmark Database.\cite{CCCBD} 

Harmonic phonon dispersions were evaluated using the finite-difference method with a \SI{0.01}{\angstrom} step-size as implemented in Phonopy.\cite{togo2015first} The forces were evaluated using the PBEsol functional with a convergence criteria of \SI{1e-6}{\eV\per\angstrom}.  Harmonic phonon theory for the calculation of free energies fails when there are phonon modes with imaginary frequency, as the contribution to the vibrational partition function becomes ill defined. In this case, anharmonic contributions need to be considered. In this study all materials which we predict to be thermodynamically stable  are also kinetically stable (ie, no phonon modes with imaginary frequency). The Supplementary Information contains phonon bandstructures for the binary materials included in this study (Figures S37 to S45).


Predictions for phase stability are highly sensitive to the calculated total energies from DFT. 
To allow comparison between layered and three-dimensional Ba-Zr-S phases with diverse bonding we evaluated the total electronic energies using the meta-GGA SCAN functional.\cite{sun2015strongly}  
A comparison against results generated using the hybrid HSE06\cite{krukau2006influence} and GGA PBEsol\cite{perdew2008restoring} functionals are provided in the Supplementary Information. 
For all total energy calculations the densities and forces were converged to \SI{e-7}{\eV\per\angstrom} and \SI{e-6}{\eV\per\angstrom} respectively.
All other inputs were set to the default value within FHI-aims. The Supplementary Information contains electronic bandstructures (Figures S30 to S36, at HSE06 and PBEsol levels of theory) for \ce{BaZrS3} and the binary materials included in this study.

\section{Results}

\subsection{\ce{BaZrS3} formation from solid materials}

\begin{figure}
    \centering
    \includegraphics[width=16cm]{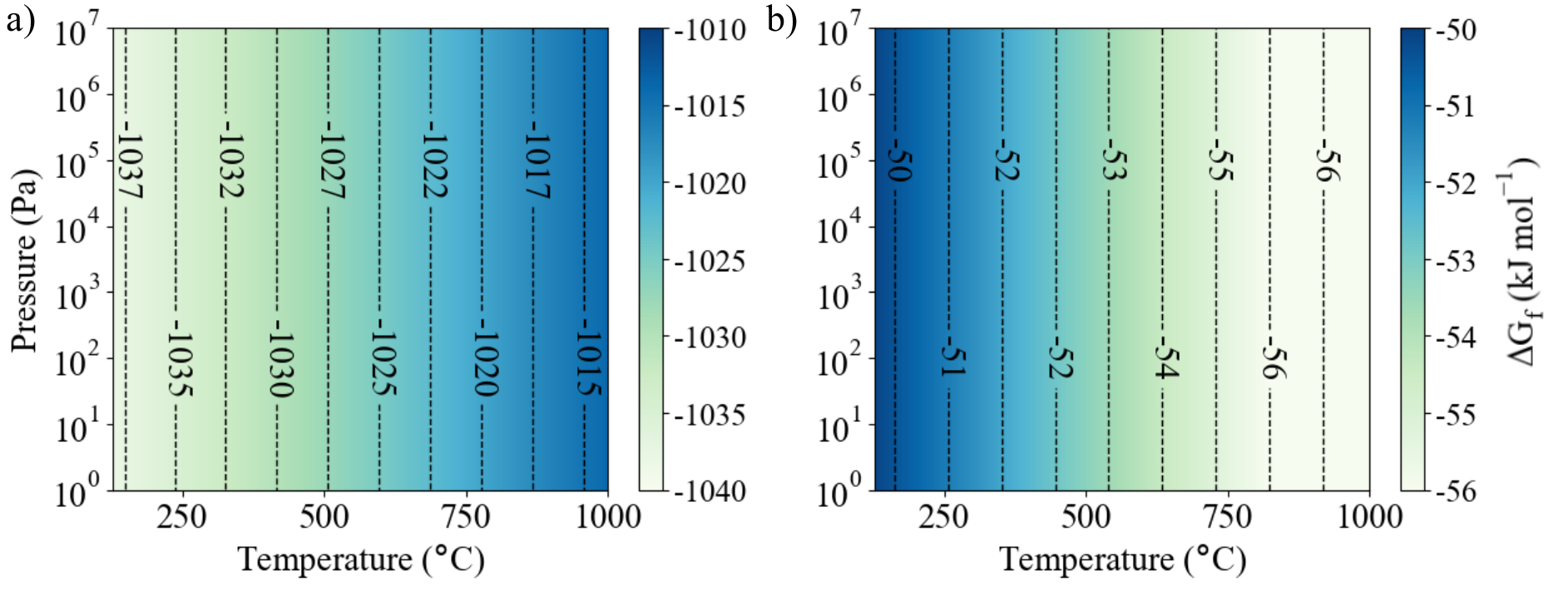}
    \caption{Gibbs free energy of formation of BaZrS$_3$ from a) its constituent elements (\ref{eq:solid_reaction_BZS}); b) binary precursors in their formal oxidation states (\ref{ep:BaS_ZrS2_BaZrS3}). As there is no sulfur-gas component, the pressure results from an inert gas or mechanical force.}
    \label{fig:solid_reaction_BZS}
\end{figure}

First, we consider formation of perovskite from its constituent elements in standard conditions, T=\SI{298}\kelvin ~and P=\SI{1E5}\pascal. At this temperature all precursors take solid form, including sulfur in the $\alpha$-phase:
\begin{chemequations}
\begin{equation}\label{eq:solid_reaction_BZS}
    \rm{Ba} + Zr + 3S(s) \rightarrow BaZrS_3.
\end{equation}
\end{chemequations}
We predict a formation enthalpy of \SI{-2.18}{\eV}/atom.
We show that reaction \ref{eq:solid_reaction_BZS} is thermodynamically feasible over a range of temperatures and pressures in Figure \ref{fig:solid_reaction_BZS}a. 
Due to the absence of a gas phase and assumed incompressibility of each solid, the Gibbs free energy of perovskite formation, $\Delta G_\mathrm{f}$, is weakly dependent on pressure. 
We conclude that across a wide range of synthesis conditions \ce{BaZrS3} is stable with respect to its elemental components.

To consider more experimentally relevant precursors we model Ba and Zr in their formal oxidation states of +2 and +4, respectively:
\begin{chemequations}
     \begin{equation}\label{ep:BaS_ZrS2_BaZrS3}
    \rm{BaS} + ZrS_2 \rightarrow BaZrS_3
\end{equation}
\end{chemequations}
Our results show that there is a thermodynamic driving force towards formation of \ce{BaZrS3} across the full temperature and pressure range (Figure \ref{fig:solid_reaction_BZS}b).
The binary precursors are more stable than the elemental species, leading to a reduction in the net energy released.
As the temperature increases \ce{BaZrS3} becomes increasingly stable against degradation into the binary phases. This follows from a relatively flat phonon dispersion for \ce{BaZrS3} at frequencies beneath \SI{2.5}{\tera\hertz}, leading to an increased density of states and vibrational entropy (Figures S5 to S7).\cite{Fultz1995phonon}

\subsection{Reaction thermodynamics in sulfur gas}

We have demonstrated that it is thermodynamically favourable to \ce{BaZrS3} from \ce{BaS} and \ce{ZrS2}, but there are severe kinetic limitations which limit this process at moderate temperatures ($\sim$\SI{600}{\celsius}).
Increasing synthesis temperature overcomes the kinetic barrier, with crystalline material typically grown at \SI{800}{\celsius} and above.\cite{Hahn1957unter,comparotto2020chalcogenide}
Sulfur partial pressure is an additional parameter which has been tuned to accelerate the reaction dynamics of \ce{BaZrS3}.
In particular, over-stoichiometric amounts of sulfur are shown to promote formation of perovskite at growth temperatures suitable for thin-film synthesis ($<$\SI{600}{\celsius}).\cite{wang2001synthesis,yu2021chalcogenide,turnley2022solution,freund2022fabrication,vincent2023liquid,yang2023low}

A common synthesis approach is to start with sulfur in the solid state ($\alpha$-sulfur), followed by heat treatment in a sealed ampule. 
During this annealing step the sulfur sublimates to form sulfur gas. 
Assuming the gas is in equilibrium, the ideal gas law can be applied to determine the sulfur partial pressure from annealing temperature, ampule volume and sulfur amount.
Another common assumption made in this calculation is that the sulfur gas is composed of a single allotrope (no allotrope mixing).
However in actuality gaseous sulfur forms a range of open and closed S$_n$ species with equilibrium allotrope ratios that are highly sensitive to temperature and pressure.\cite{Dobbie1919The,jackson2016universal}
Furthermore, annealing may occur in non-equilibrium conditions resulting from the experimental setup, such as gas flow, gas escape or temperature gradients.\cite{Ren2017Evolution}
Experimental insight into reaction conditions required for \ce{BaZrS3} synthesis is limited by the challenges associated with in-situ monitoring of sulfur gas partial pressure during material synthesis.

For a system in equilibrium, quantum chemical modelling is deemed to give the most accurate understanding of sulfur gas constitution.\cite{Steudel2003speciation} 
Jackson et al. use a global structure search and first-principles lattice dynamics to identify the equilibrium ratios of 13 low-energy sulfur gas allotropes. \cite{jackson2016universal}
They conclude that the dominant components are \ce{S2} and \ce{S8}.
The trend is for high-$n$ ($n$=8) species to dominate at low temperatures and high pressures, and low-$n$ ($n$=2) species to dominate at high temperatures and low pressures. This is in agreement with experimental measurements.\cite{Dobbie1919The,Steudel2003speciation,jackson2016universal}
The transition temperature at which the dominant phase changes ($\mu_{S_2} = \mu_{S_8}$) is highly 
pressure- dependent, with higher pressures corresponding to higher transition temperatures. 
In addition higher pressures correspond to a wider temperature range in which contributions from the cyclic allotropes \ce{S4}--\ce{S7} are significant.\cite{jackson2016universal}

In this work, we consider three sulfur gas compositions; i) single-component \ce{S2} (the smallest allotrope); ii) single-component \ce{S8} (the largest allotrope); iii) an equilibrium mixture of the \ce{S2}--\ce{S8} allotropes (13 species in total), denoted \ce{S_{mix}}. 
The enthalpy change when converting between sulfur allotropes is largest for \ce{S2} and \ce{S8} at moderate temperatures,\cite{Berkowitz2004Equilibrium} so that these systems reflect the extremes of behaviour that might be expected for (out of equilibrium) single-allotrope systems.
\ce{S_{mix}} corresponds to a gas is in equilibrium, which we assume here provides the most accurate model for comparison against experiment.

To demonstrate our approach, we consider perovskite formation from elemental compounds with sulfur in the gas phase. 
We consider an atmosphere with sulfur vapour only, so that the sulfur partial pressure is equal to total pressure.
For the higher-pressure, lower-temperature regime the \ce{S8} allotrope is more stable than the \ce{S2} allotrope, with $\mu_{S_8} < \mu_{S_2}$ where $\mu$ is the chemical potential of a single atom (Figure 3a). As such, in this regime we expect perovskite formation in equilibrium with the \ce{S8} allotrope to be the more relevant reaction (Figure 3b):
\begin{chemequations}
\begin{equation}\label{eq:S8_reaction_BZS}
    \rm{Ba} + Zr + \frac{3}{8}S_8(g) \rightarrow BaZrS_3.
\end{equation}
\end{chemequations}

For the lower-pressure, higher-temperature regime where $\mu_{S_2} < \mu_{S_8}$ the equilibrium with the \ce{S2} allotrope is more relevant (Figure 3c):
\begin{chemequations}
\begin{equation}\label{eq:S2_reaction_BZS}
    \rm{Ba} + Zr + \frac{3}{2}S_2(g) \rightarrow BaZrS_3.
\end{equation}
\end{chemequations}

\begin{figure*}
    \centering
    \includegraphics[width=16cm]{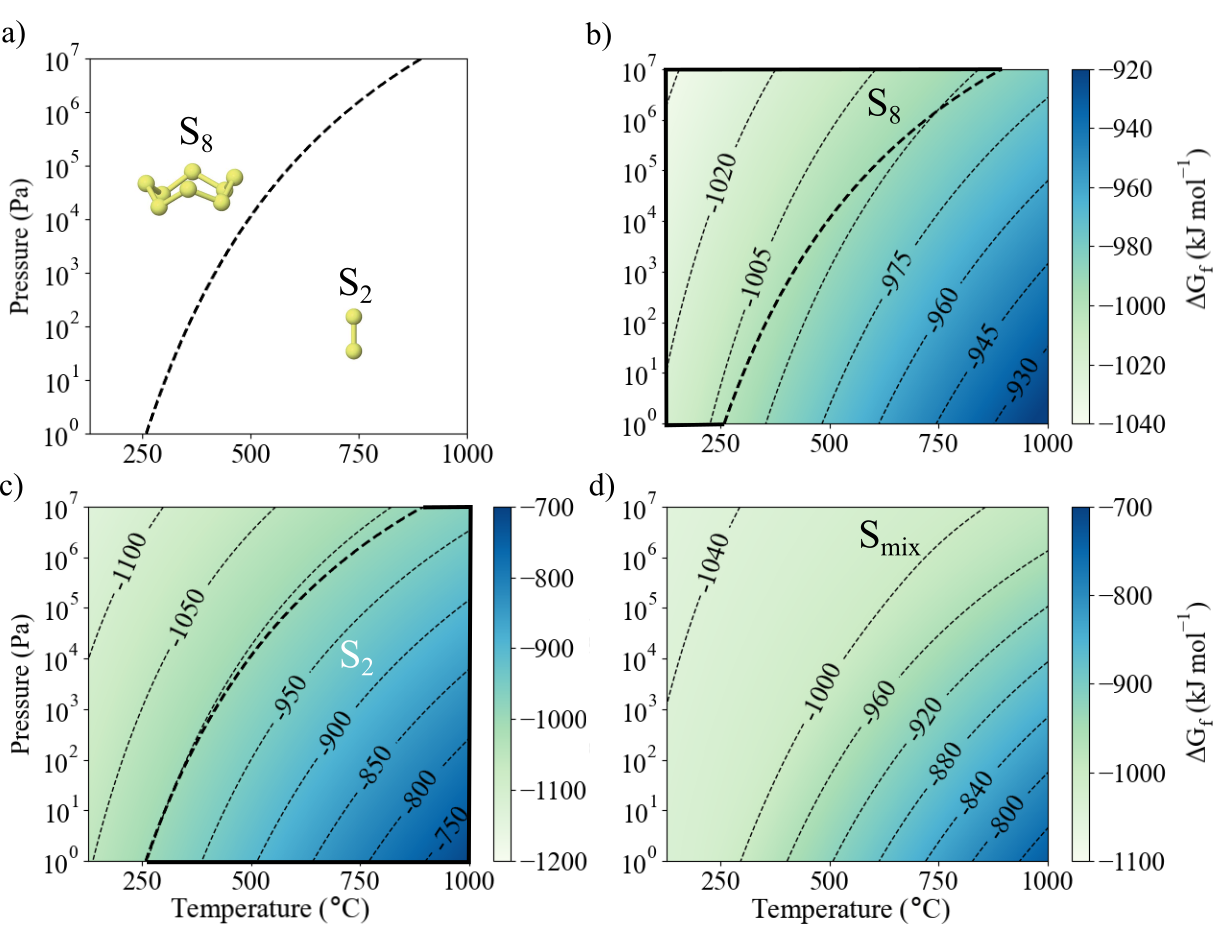}
    \caption{a) The coexistence curve for single allotrope \ce{S2} vapour and single allotrope \ce{S8} vapour. The dashed line indicates the coexistence curve where the chemical potential $\mu$ of the sulfur gas allotropes on a per-atom basis are equal, $\mu_{{S_8}}=\mu_{{S_2}}$. At lower temperatures and higher pressures \ce{S8} dominates, at higher temperatures and lower pressures \ce{S2} dominates; b) Gibbs free energy of \ce{BaZrS3} formation from solid Ba, Zr and \ce{S8} (R3 in main text). The dashed and heavy solid lines indicate the region where the \ce{S8} allotrope dominates; c) Gibbs free energy of \ce{BaZrS3}formation from solid Ba, Zr and \ce{S2} (R4). The dashed and heavy solid lines indicates the region where the \ce{S2} allotrope dominates;
    (d) Gibbs free energy of \ce{BaZrS3} formation from solid Ba, Zr and S$_\mathrm{mix}$ (R5). The chemical potential of S$_\mathrm{mix}$ incorporates the effects of \ce{S2} to \ce{S8} allotrope mixing.\cite{jackson2016universal}}
    \label{fig:Fig3}
\end{figure*}

The range of experimental conditions used for \ce{BaZrS3} (precursor) synthesis is coincident with the coexistence curve in Figure \ref{fig:Fig3}a and so we expect allotrope mixing to impact on reaction thermodynamics. Furthermore we require a model that is valid across the temperature and pressure ranges considered for experiment. As such, we consider equilibrium with \ce{S_{mix}} to be the most relevant reaction for comparison against experiment.
\begin{chemequations}
\begin{equation}\label{eq:Sn_reaction_BZS}
    \rm{Ba} + Zr + 3\ce{S_{mix}}(g) \rightarrow BaZrS_3.
\end{equation}
\end{chemequations}

For all sulfur gas compositions we find that the formation of \ce{BaZrS3} is thermodynamically favoured. 
As expected, increased sulfur partial pressure stabilises the perovskite, as the gas species favours entry into a low-pressure environment. 
The temperature dependence is also more acute compared to formation from $\alpha$-sulfur. This follows from the free translational and rotational motion of a gas molecule leading to higher entropy.

We find that sulfur allotrope can have a noticeable impact on $\Delta G_\mathrm{f}$.
For example, under typical processing conditions (T=\SI{500}{\celsius}, P=\SI{1E2}{\pascal}) the predicted $\Delta G_\mathrm{f}$ is \SI{34}{\kilo\joule\per\mole} less when forming perovskite in equilibrium with \ce{S8} (Reaction \ref{eq:S8_reaction_BZS}) compared to an equilibrium with \ce{S2} (Reaction \ref{eq:S2_reaction_BZS}). 
This is consistent with \ce{S2} being the stable allotrope at T=\SI{500}{\celsius}, P=\SI{1E2}{\pascal}.

To understand the impact of allotrope mixing we compare the single species models (Figures 3b and 3c) against 
the mixed-allotrope model (Figure 3d). 
As expected, in the lower-temperature and higher-pressure regions where the \ce{S8} allotrope dominates, there is a close agreement between $\Delta G_\mathrm{f}$ calculated for a reaction with the single allotrope \ce{S8} and the mixed allotrope \ce{S_{mix}}. 
In the higher-temperature and lower-pressure region where the \ce{S2} allotrope dominates there is closer agreement between \ce{S2} and \ce{S_{mix}} than \ce{S8} and \ce{S_{mix}}.
The mixed- and single-allotrope predictions deviate most strongly in the region where the sulfur model predicts significant mixing between allotropes, which encompasses the region typically targeted for \ce{BaZrS3} formation (Figures S8 and S9).
This indicates that the extent of allotrope mixing may have an impact on the thermodynamic feasibility of reaction processes.




\subsection{Formation of sulfur-rich binary precursors}

The critical temperature for accelerated perovskite growth coincides with the low-melting point of the intermediate phase \ce{BaS3} at \SI{554}{\celsius}.\cite{binaryalloy2016} 
Following this it is proposed that \ce{BaS3} acts as a liquid flux which overcomes the kinetic barriers associated with solid-state precursors.\cite{sopiha2022chalcogenide,freund2022fabrication,vincent2023liquid,yang2023low}
An additional source of sulfur during annealing gives access to this sulfur-rich binary phases.
Although experimental studies have started to explore the conditions under which \ce{BaS3} is formed,\cite{yang2023low,vincent2023liquid,freund2022fabrication}
the upper- and lower- limits of sulfur partial pressure is currently undefined.
\ce{ZrS3} formation has also been reported during \ce{BaZrS3} synthesis, with
several studies suggesting that forming \ce{ZrS3} as a reaction intermediate can hinder the formation of perovskite.\cite{wang2001synthesis,yang2023low}

To predict which solid binary materials are most stable at a particular temperature and sulfur partial pressure we assume that precursors are in the (Ba,Zr)-S system and that the reaction occurs in equilibrium with a mixed-allotrope sulfur atmosphere. For example, to compare the stability of \ce{BaS} and \ce{BaS2} we calculate $\Delta G_\mathrm{f}(T,P)$ for the following reaction:
\begin{chemequations}
     \begin{equation}\label{eq:BaS-Smix--BaS2}
     \ce{BaS} + \ce{S_{mix}}\mathrm{(g)}  \rightarrow \ce{BaS2} 
     \end{equation}
\end{chemequations} 
If $\Delta G_\mathrm{f}$ for \ref{eq:BaS-Smix--BaS2} is negative we predict that \ce{BaS2} will form in preference to \ce{BaS} at that particular temperature and pressure.
We conduct this analysis for all materials reported to be within \SI{0.5}{\electronvolt} of the convex hull when calculated using ground-state DFT (see the Supplementary Information for more details on the database search). For Ba-S this corresponds to calculating the relative stability of \ce{BaS}, \ce{Ba2S3}, \ce{BaS2} and \ce{BaS3}. For Zr-S we consider \ce{ZrS}, \ce{Zr3S4}, \ce{ZrS2} and \ce{ZrS3}. 
 
\begin{figure*}
    \centering
        \includegraphics[width=16cm]{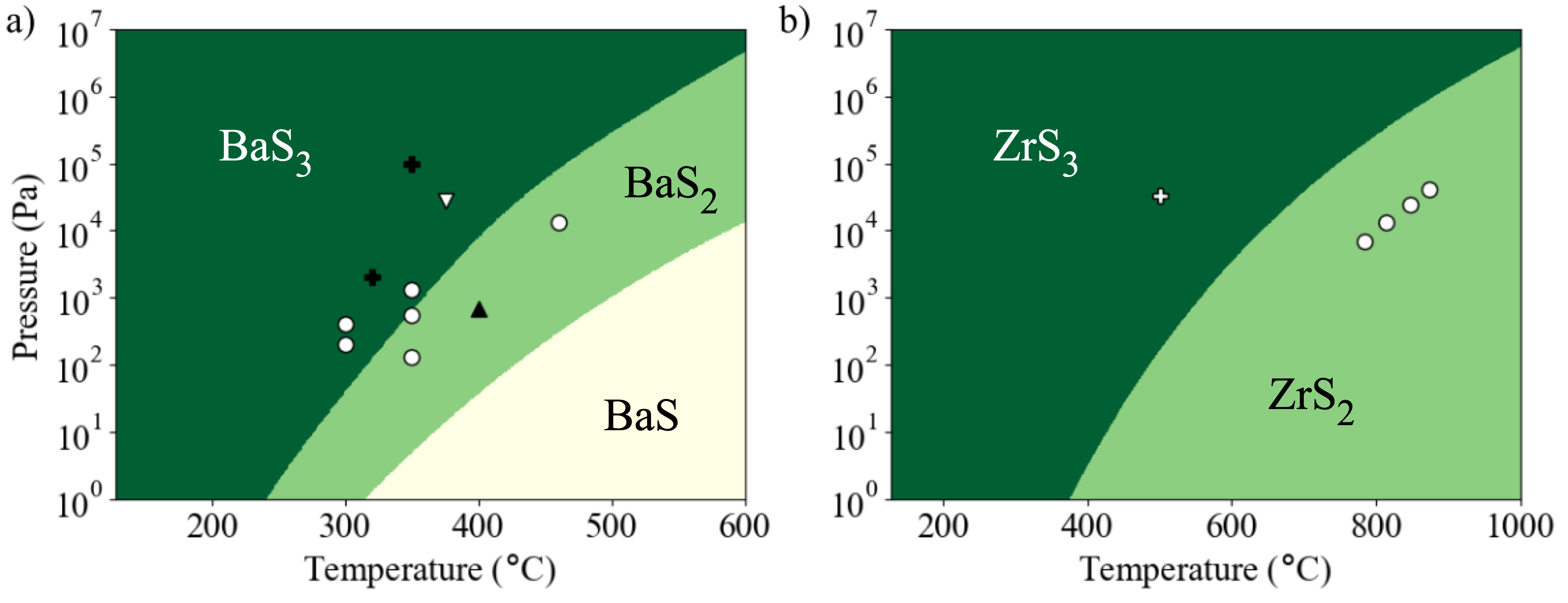}
    \caption{Diagram displaying the most stable compounds when in equilibrium with sulfur vapour a) Ba-S materials. The scatter points are taken from published experimental data. White filled shapes correspond to \ce{BaS3} formation: circles from Reference \citenum{freund2022fabrication} and triangle from Reference \citenum{vincent2023liquid}. The black filled triangle corresponds to \ce{BaS2} and the black filled crosses correspond to \ce{BaS}, both from Reference \citenum{freund2022fabrication}. Note that we do no consider the liquid phases of \ce{BaS_x} which are known to form above \SI{554}{\celsius};\cite{binaryalloy2016} b) Zr-S materials. The scatter points are taken from published experimental data with the cross from Reference \citenum{vincent2023liquid} denoting \ce{ZrS3} formation. The circles from Reference \citenum{Haraldsen1963On} denote the points at which \ce{ZrS3} decomposes to \ce{ZrS2}.}
    \label{fig:Ba-S}
\end{figure*}

The most stable materials are displayed as a function of temperature and pressure in Figure 4.
For the Ba-S system we find that the sulfur partial pressure has a marked impact on stability at moderate temperatures. 
For example at \SI{400}{\celsius} \ce{BaS2} formation is favourable between \SIrange{4E1}{1E4}{\pascal}, with \ce{BaS} formed below this range and \ce{BaS3} is formed above.
As the temperature increases, the partial pressures required for sulfur-rich compounds to form also increases.
\ce{Ba2S3} is predicted to be unstable across the temperature and pressure ranges considered in this analysis.

There is limited data available for \ce{BaS3} formation in an evacuated ampule with a sulfur source.
Yang et al. anneal \ce{BaS} at \SI{400}{\celsius} and find conversion into \ce{BaS3}.\cite{yang2023low} 
Vincent et al. use a similar approach and find conversion to \ce{BaS3} at \SI{375}{\celsius} and at an estimated sulfur partial pressure of \SI{0.28E5}{\pascal}.\cite{vincent2023liquid} 
This measurement is within the range of pressures we predict for \ce{BaS3} stability. 
After removing the sulfur source from the ampule, Vincent et al. observe decomposition into \ce{BaS2} at \SI{575}{\celsius}. 
This indicates that \ce{BaS3} is not stable at moderate temperatures and reduced partial pressure, in line with our predictions.

Freund et al. anneal \ce{BaS} in a sulfur-nitrogen atmosphere across a range of temperatures (\SIrange{250}{460}{\celsius}) and total pressures (\SIrange{1.3E2}{1E5}{\pascal}).\cite{freund2022fabrication}
They report \ce{BaS3} formation at pressures within and below our predicated range.
They also find that for a given temperature if the pressure was too high then no reaction occured and \ce{BaS} remained. This is contrary to equilibrium gas behaviour, suggesting a change in reaction conditions or onset of a competing reaction, which is not captured in our model. 
Comparotto et al. report synthesis of \ce{BaZrS3} across a range of sulfur partial pressures.\cite{comparotto2020chalcogenide} They report improved crystallinity at higher sulfur partial pressures and propose that this may be a consequence of \ce{BaS3} liquid-flux formation. However the upper bound for pressure is \SI{5}{\pascal} at \SI{590}{\celsius}, which is significantly below the lower bound we predict for \ce{BaS3} formation. We note that as we do not consider \ce{BaS3} in the liquid phase it is likely that this is stabilised at this temperature for lower partial pressures.

For the Zr-S system in Figure 4b our model predicts that \ce{ZrS2} and \ce{ZrS3} are the only thermodynamically stable materials.
Experimental data for the decomposition of \ce{ZrS3} into \ce{ZrS2} and sulfur gas lies within one order of magnitude of our model (Figure 4b).\cite{Haraldsen1963On}  
The scaling relationship between decomposition temperature and pressure is remarkably close to what we predict, with higher temperatures corresponding to higher pressures. 
Vincent et al. report conversion of \ce{ZrH2} precursor to \ce{ZrS3} at \SI{500}{\celsius} and sulfur partial pressure of \SI{3.3E4}{\pascal}, which lies in the predicted region of \ce{ZrS3} formation.\cite{vincent2023liquid}

We find that \ce{ZrS3} is formed across a wider range of temperatures and pressures than \ce{BaS3}. For example, at \SI{500}{\celsius} \ce{ZrS3} is formed at \SI{2E2}{\pascal}, compared to \SI{3E5}{\pascal} for \ce{BaS3}.
The consequence of this is that in certain conditions \ce{BaS2} and \ce{ZrS3} will co-exist. 
Yang et al. report that after combining \ce{BaS3} and \ce{ZrS2} powders (with no sulfur excess) in a vacuum sealed ampule a reaction is initiated to form \ce{BaS2} and \ce{ZrS3} at \SI{500}{\celsius}. This is in agreement with our prediction that $\Delta G_\mathrm{f}$ is \SI{-15.8}{\kilo\joule\per\mole} for the corresponding reaction
  \begin{chemequations}
     \begin{equation}\label{S_transfer}
     \ce{BaS3} + \ce{ZrS2} \rightarrow \ce{BaS2} + \ce{ZrS3},
    \end{equation}
 \end{chemequations}
with sulfur vapour proposed as a reaction intermediate (see R1 and R2 in the Supplementary Information).

We find that the positions of the coexistence curves(s) are sensitive to sulfur vapour composition, with \ce{S8} hindering the formation of \ce{BaS3} across temperature and pressure ranges used for synthesis (Figures S11 and S12 in the Supplementary Information). 
This suggests that sulfur allotrope may be a significant factor in the thermodynamics of \ce{BaS3} formation; equilibrium and non-equilibrium sulfur vapours may lead to different chemical behaviour.

In conclusion, we find that there is high sensitivity to the sulfur partial pressure, and this must be carefully controlled to access the desired binary intermediates, and avoid the undesired. 
At \SI{500}{\celsius} a sulfur partial pressure above \SI{3E5}{\pascal} is required to form \ce{BaS3}. 
At this point it also becomes thermodynamically favourable to form \ce{ZrS3}, which is reported to hinder \ce{BaZrS3} formation. 
As such, our model supports the recently reported strategy of using a liquid \ce{BaS3} precursor rather than an annealing step with an additional sulfur source.\cite{yang2023low}

\subsection{\ce{BaZrS3} stability against degradation into sulfur rich binaries}
\begin{figure}
    \centering
        \includegraphics[width=16cm]{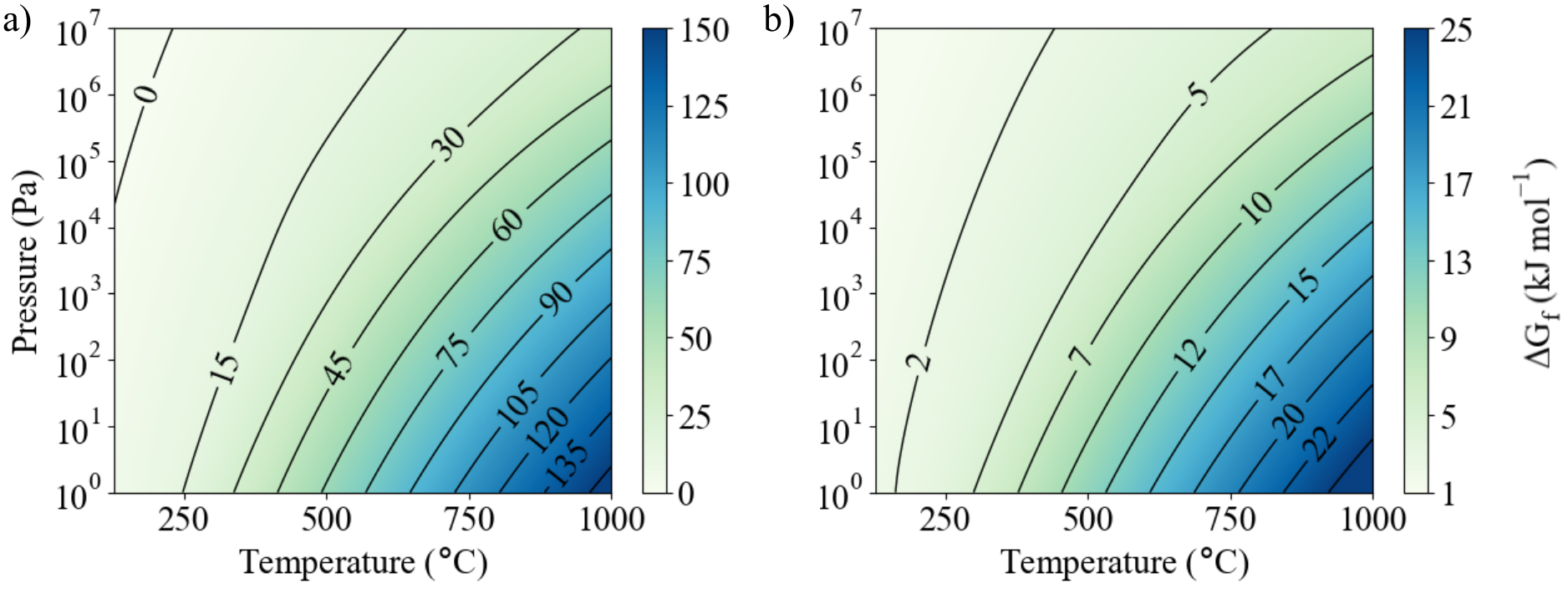}
    \caption{Gibbs free energy of \ce{BaZrS3} degradation when in equilibrium with a mixed-allotrope sulfur gas a) Degradation into \ce{ZrS3} and \ce{BaS3} (Reaction \ref{eq:BZS-Sn--BaS3-ZrS3}); b) Degradation into \ce{ZrS3} and \ce{BaS2} (Reaction \ref{eq:BZS-Sn--BaS2-ZrS3})}
    \label{fig:BaS3_ZrS3_BZS}
\end{figure}
Several experimental studies have established that if the partial pressure of sulfur is too high during annealing, \ce{ZrS3} growth is favored at the expense of \ce{BaZrS3} formation.
Two mechanisms have been proposed for this.\cite{sopiha2022chalcogenide} High partial pressures may cause \ce{BaZrS3} to become unstable:
  \begin{chemequations}
     \begin{equation}\label{eq:BZS-Sn--BaS3-ZrS3}
     \ce{BaZrS3} + 3\ce{S_{mix}}\mathrm{(g)} \rightarrow \ce{ZrS3} + \ce{BaS3} 
    \end{equation}
      \begin{equation}\label{eq:BZS-Sn--BaS2-ZrS3}
     \ce{BaZrS3} + 2\ce{S_{mix}}\mathrm{(g)}  \rightarrow \ce{ZrS3}  + \ce{BaS2}
    \end{equation}
 \end{chemequations}
Or, alternatively, \ce{ZrS2} forms \ce{ZrS3} in a sulfur rich atmosphere:
  \begin{chemequations}
     \begin{equation}\label{}
     \ce{ZrS2} + \ce{S_{mix}}\mathrm{(g)}  \rightarrow \ce{ZrS3}, 
    \end{equation}
 \end{chemequations}
with \ce{ZrS3} mass transport limiting perovskite formation. 

We have demonstrated in the previous section that \ce{ZrS3} can form across a wide range of temperatures in a sulfur rich atmosphere. In Figure 5 we consider reactions \ref{eq:BZS-Sn--BaS3-ZrS3} and \ref{eq:BZS-Sn--BaS2-ZrS3}. We find that \ce{BaZrS3} in equilibrium with sulfur gas is stable against degradation into \ce{ZrS3}.
This result is supported by a recent study demonstrating a complete conversion of \ce{BaS3} and \ce{ZrS3} into \ce{BaZrS3} for a reaction time of three days.\cite{yang2023low} 
We note that whilst \ce{BaZrS3} might be stable with respect to binary materials in a sulfur-rich environment, the increased chemical potential of sulfur may lead to other unwanted effects - for example, the formation sulfur interstitial defects. Of particular concern is the S$_\mathrm{Zr}$ antisite substitution which is predicted to have a low formation energy and electronic levels within the \ce{BaZrS3} band gap.\cite{Meng2016Alloying}

In the Supplementary Information we discuss the analysis of perovskite degradation into all binary and elemental precursors within \SI{0.5}{\eV} of the convex hull. In all cases, \ce{BaZrS3} remains stable. When \ce{BaZrS3} is formed in an atmosphere with unstable single-allotrope vapour \ce{S2} (Figure S15) we predict degradation only at pressures above the saturation vapour pressure for sulfur, where our ideal gas model is no longer valid.\cite{West1929the} 

\section{Conclusions}

In this work we have considered reported phases in the Ba-Zr-S system within \SI{0.5}{\electronvolt} of the convex hull.
Our results demonstrate the thermodynamic feasibility of forming perovskite at moderate temperatures; we predict that up to a temperature of \SI{1000}{\celsius} \ce{BaZrS3} is stable against decomposition into solid binary or elemental competing phases. Importantly, this is true when annealed in mixed-allotrope (equilibrium) sulfur vapour at high partial pressures. 

Our model predicts that several sulfr-rich \ce{BaS_x} and \ce{ZrS_x} materials can be formed when the binary systems are annealed in sulfur vapour.
A general trend is that to form a more sulfur-rich species the temperature must be reduced, or the sulfur partial pressure increased.
At \SI{500}{\celsius} a sulfur partial pressure above \SI{3E5}{\pascal} is required to form \ce{BaS3}.
At any temperature and pressure where \ce{BaS3} is formed, \ce{ZrS3} is also predicted to form. 
For this reason, our results support the idea of using \ce{BaS3} and \ce{ZrS2} precursors for fast perovskite synthesis, before kinetically-limiting \ce{ZrS3} can be formed.

We have focused our analysis on the most commonly reported competing phases or reaction intermediates, 
and emphasize that thermodynamic free energies for all possible product and reactant combinations reported in this study can be predicted using our computational notebook\cite{PaperRepo} and ThermoPot software.\cite{ThermoPot}
ThermoPot is material-agnostic so can be used to post-process first-principles calculations for other systems; the same approach outlined here could be applied to alloy chalcogenide perovskites, for example. Extension to other gas species is possible where there is published thermochemical data.

To widen the material space for this system or incorporate the effects of disorder recent advances in computational modelling could be applied. For example, thermodynamic integration techniques to incorporate liquid \ce{BaS_x} phases, or statistical methods to quantify the configurational entropy term associated with Ruddlesden-Popper phase inter-mixing. Oxide materials might also play a role in \ce{BaZrS3} formation, with \ce{ZrO2} formation potentially acting as a Zr `sink'. 

Compared to experimental measurements, our model over-estimates the partial pressure at which \ce{BaS3} and \ce{ZrS3} begins to form. 
This motivates further studies to close the gap between theory and experiment.
On the theory side, the ab-initio model for lattice dynamics could be generalized to include the effects of thermal expansion (using the quasi-harmonic approximation) or higher-order lattice anharmonicity (using a temperature dependent effective potential, for example).
For experiment, improved experimental measurement and control of temperature and sulfur partial pressure, with analysis that accounts for allotrope mixing, would provide more accurate measurements for comparison.

A key result is that the constitution of sulfur vapour has a significant impact on reaction thermodynamics. 
It is difficult to maintain a constant temperature and sulfur partial pressure during chalcogenide material synthesis and---as has been seen for \ce{CZTS}\cite{Zhang2019Non}---the proportion of sulfur allotropes may fluctuate depending on sample position, dwell time, and the particulars of the experimental setup.\cite{Ren2017Evolution} 
As a result there could be challenges for experimental reproducibility, with fluctuations in the rate and feasibility of perovskite formation for comparable setups.
This insight can be transferred to other sulfide materials with phase equilibria in temperature and pressure regions with significant allotrope mixing, and motivates further work in the characterisation and control of the sulfur vapour.


\begin{acknowledgement}

  We thank Jonathan Scragg, Kostiantyn Sopiha and Corrado Comparotto for useful discussions related to this work. P.K. acknowledges support from the UK Engineering and Physical Sciences Research Council (EPSRC) CDT in Renewable Energy Northeast Universities (ReNU) for funding through EPSRC Grant EP/S023836/1. This work used the Oswald High-Performance Computing facility operated by Northumbria University (UK). Via our membership of the UK’s HEC Materials Chemistry Consortium, which is  funded by EPSRC (EP/X035859), this work used the ARCHER2 UK National Supercomputing Service \href{https://www.archer2.ac.uk/}.
  
\end{acknowledgement}

\begin{suppinfo}

Phonon dispersions, electronic band structures, additional computational details and analysis are provided in the Supplementary Information.

Calculation input and output files (including the relaxed geometries, total energies, and phonon data) are available in both project-specific\cite{PaperRepo} and NoMaD\cite{NOMAD} repositories.
Computational notebooks to reproduce the results reported here are also available in the project-specific repository.

\end{suppinfo}


\bibliography{main} 

\end{document}




\newpage
\section{Database search}\label{sec:identification_of_competing_phases}
From the Materials Project v2022.10.28,\cite{Jain2013Commentary} we query for the binary Ba-S and Zr-S systems and the ternary Ba-Z-S systems. We select the compounds that i) have been experimentally reported on the ICSD;\cite{zagorac2019recent} ii) lie within \SI{0.5}{\electronvolt} of the convex hull (as calculated using the default GGA functional); iii) have primitive cells containing less than 15 atoms.

\begin{table}[h!]
    \centering
    \begin{tabular}{llll}
        Formula & Space Group & Materials Project ID & ICSD no. \\
        \hline
        BaS & $Fm\Bar{3}m$ & mp-1500 & 2004\\
        BaS$_2$ & $C2/c$ & mp-684 & 42134\\
        BaS$_3$ & $P\Bar{4}2_1m$ & mp-239 & 23637\\
        BaS$_3$ & $P2_12_12$ & mp-556296 & 26765\\
        Ba$_2$S$_3$ & $I4_1md$ & mp-28978 & 70058\\
        ZrS & $Fm\Bar{3}m$ & mp-1925 & 52224\\
        ZrS & $P4/nmm$ & mp-7859 & 24754\\
        ZrS$_2$ & $P\Bar{3}m1$ & mp-1186 & 76037\\
        ZrS$_3$ & $P2_1/m$ & mp-9921 & 42073\\
        Zr$_3$S$_4$ & $Fd\Bar{3}m$ & mp-1103820 & 108734\\
        BaZrS$_3$ & $Pnma$ & mp-540771 & 23288\\
        Ba$_2$ZrS$_4$ & $I4/mmm$ & mp-3813 & 69853\\
        Ba$_3$Zr$_2$S$_7$ & $P4_2/mnm$ & mp-8570 & 264213\\
        Ba$_3$Zr$_2$S$_7$ & $I4/mmm$ & mp-9179 & 75241\\
        Ba$_4$Zr$_3$S$_{10}$ & $I4/mmm$ & mp-14883 & 72656\\
        \hline
        
    \end{tabular}
    \caption{Materials selected through our database search}
    \label{tab:dataset}
\end{table}

We have already established that \ce{BaZrS3} decomposes to the Ruddlesden-Popper (RP) phases and \ce{ZrS2} at high temperatures only ($>$\SI{1200}{\kelvin}).\cite{kayastha2023high} In addition, an equilibrium with sulfur vapour will make the formation of relatively sulfur-poor RP phases less thermodynamically favourable. In this study we consider perovskite formation at moderate temperatures in sulfur vapour, so we do not include an analysis of RP-phase formation.

\begin{figure}[H]
    \centering
    \includegraphics[width=0.5\textwidth]{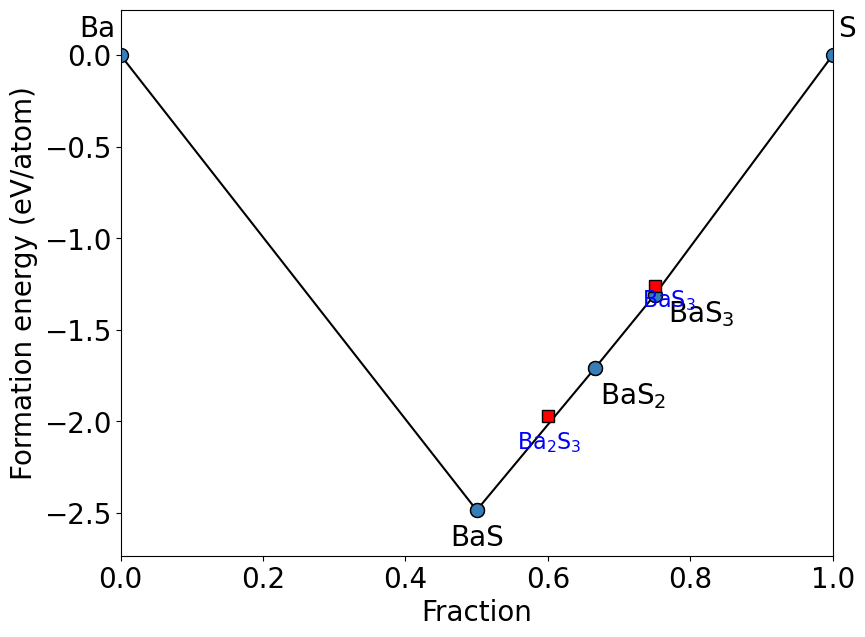}
    \caption{Convex hull at \SI{0}{\kelvin} for materials in the Ba-S system. Red points denote materials which are metastable (above the convex hull). }
    \label{fig:0_T_convex_hull_Ba-S}
\end{figure}

\begin{figure}[H]
    \centering
    \includegraphics[width=0.5\textwidth]{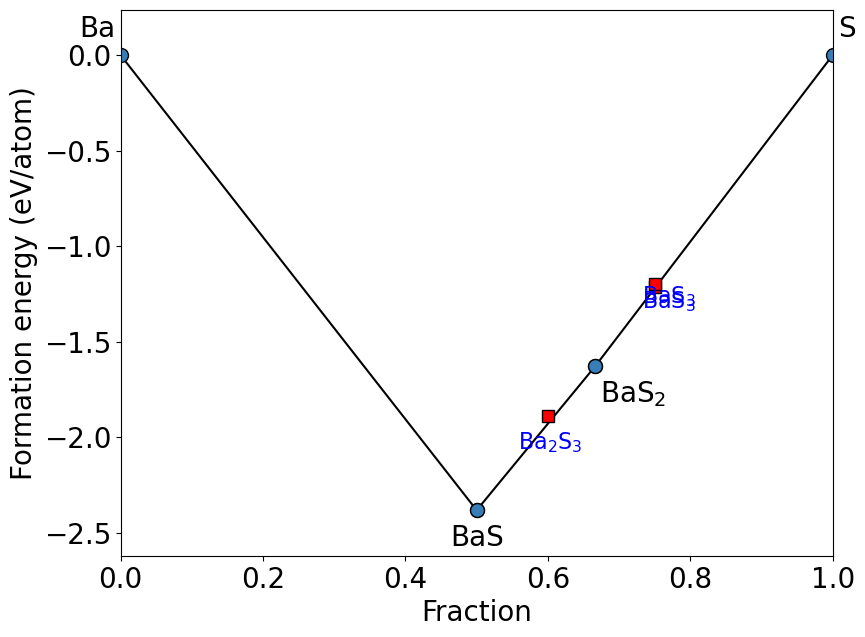}
    \caption{Convex hull at \SI{1273}{\kelvin} for materials in the Ba-S system.  Red points denote materials which are metastable (above the convex hull). }
    \label{fig:finite_T_convex_hull_Ba-S}
\end{figure}

\begin{figure}[H]
    \centering
    \includegraphics[width=0.5\textwidth]{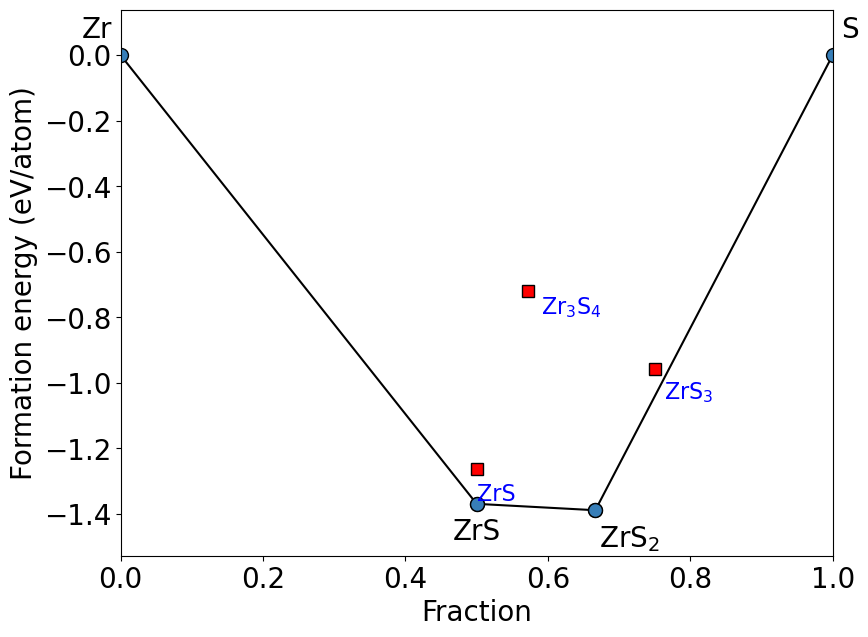}
    \caption{Convex hull at \SI{0}{\kelvin} for materials in the Zr-S system. Red points denote materials which are metastable (above the convex hull). }
    \label{fig:0_T_convex_hull_Zr-S}
\end{figure}

\begin{figure}[H]
    \centering
    \includegraphics[width=0.5\textwidth]{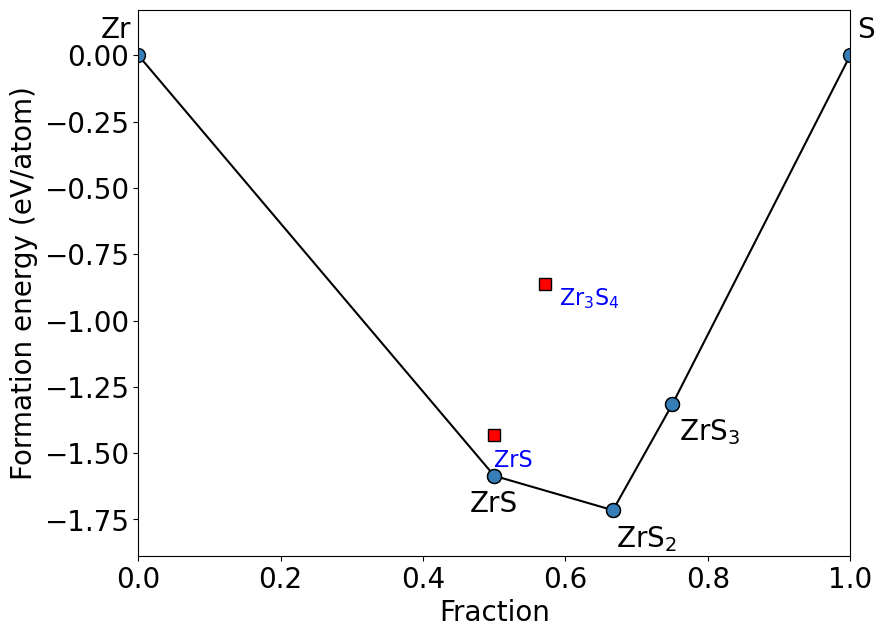}
    \caption{Convex hull at \SI{1273}{\kelvin} for materials in the Zr-S system. Red points denote materials which are metastable (above the convex hull). }
    \label{fig:finite_T_convex_hull_Zr-S}
\end{figure}

The total energies used to construct the convex hull correspond to the total electronic energies calculated using ground-state Density Functional Theory. 
A \SI{0.5}{\electronvolt} cutoff above the convex hull was chosen as this is reported to include the 90th percentile of all metastable materials reported within the Materials Project.\cite{sun2016thermodynamic,aykol2018thermodynamic} 
To assess if this is likely to capture all of the materials which might be stabilised within the temperature and pressure ranges we are considering, we evaluated the $PV$ and $TS$ terms for each binary material in Table \ref{tab:dataset} using first-principles thermodynamics, as outlined in the methods section of the main text. We used the resulting Gibbs free energies and pymatgen\cite{Ong2008Li} to plot the convex hull at \SI{1000}{\celsius}.
For incompressible solids the effect of pressure is negligible; in this case we consider $P=$\SI{1}{\pascal}. In Figures \ref{fig:0_T_convex_hull_Ba-S} to \ref{fig:finite_T_convex_hull_Zr-S} we show that for both systems the largest change in energy relative to the convex hull is less than \SI{0.5}{\electronvolt}, which suggests our criteria for the selection of ground-state materials is reasonable. All energies are calculated using the SCAN exchange-correlation functional.\cite{sun2015strongly}

\newpage

\section{Entropic contributions to the Gibbs Free Energy of formation}

\begin{figure}[H]
    \centering
    \includegraphics[width=0.8\textwidth]{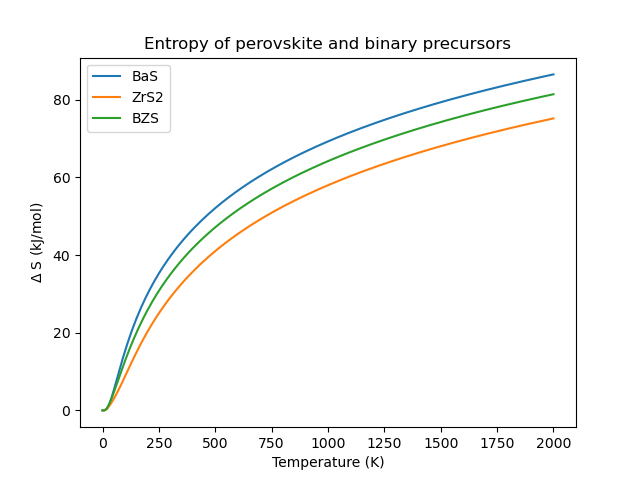}
    \caption{Entropy of \ce{BaZrS3} perovskite and binary precursors calculated using first principles lattice dynamics. BZS denotes \ce{BaZrS3}.}\label{fig:entropy1}
\end{figure}

\begin{figure}[H]
    \centering
    \includegraphics[width=0.8\textwidth]{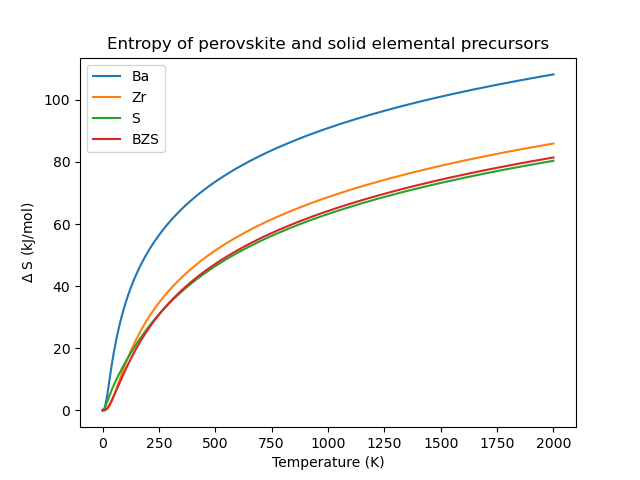}
    \caption{Entropy of \ce{BaZrS3} perovskite and solid elemental precursors calculated using first principles lattice dynamics. BZS denotes \ce{BaZrS3}.}\label{fig:entropy2}
\end{figure}

\begin{figure}[H]
    \centering
    \includegraphics[width=0.8\textwidth]{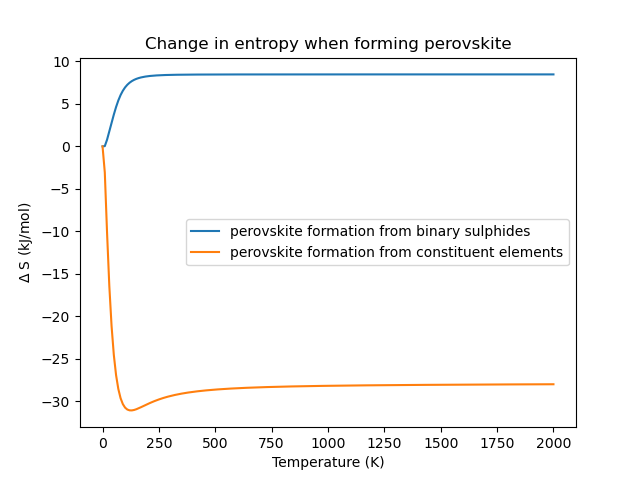}
    \caption{Change in entropy when forming \ce{BaZrS3} perovskite from solid elemental precursors or binary precursors. BZS denotes \ce{BaZrS3}.}\label{fig:entropy3}
\end{figure}

In Fig 2b of the main text we see that the perovskite product becomes increasingly stabilised with temperature for the reaction $\rm{BaS} + ZrS_2 \rightarrow BaZrS_3$. This stems from a net entropy gain during perovskite formation. This can be monitored by extracting the $S$ term from our lattice dynamics calculations; see Figure \ref{fig:entropy1} for the entropy of each component, and Figure \ref{fig:entropy3} for the change in entropy during perovskite formation. We postulate that the relatively large perovskite entropy term results from the high density of flat bands beneath \SI{2.5}{\tera\hertz} in Figure \ref{fig:P9}, which likely leads to an increased density of states at lower phonon frequencies.
In Fig 2a of the main text we see the opposite behaviour: perovskite formation from elemental compounds (Figures \ref{fig:entropy2} and \ref{fig:entropy3}) becomes less favourable with increasing temperature (a net entropy loss).

\newpage

\section{Comparison between the single-allotrope and mixed-allotrope models}

\begin{figure}[H]
    \centering
    \includegraphics[width=0.5\textwidth]{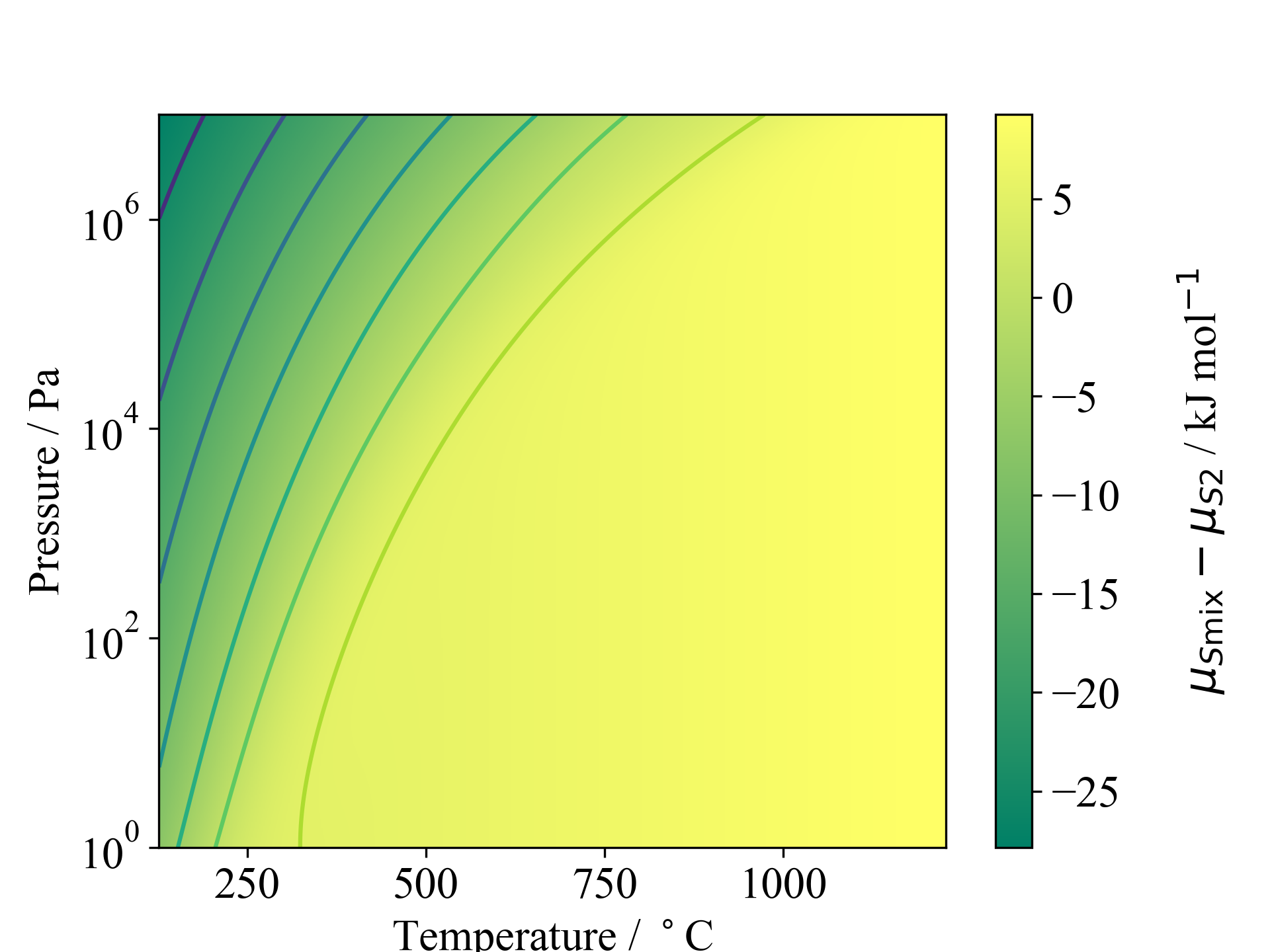}
    \caption{Comparison of the Gibbs free energy associated with a single allotrope sulfur vapour \ce{S2} and mixed allotrope S$_\mathrm{mix}$. The pressure is the partial pressure of the sulfur gas.}
    \label{fig:S2_mix}
\end{figure}

\begin{figure}[H]
    \centering
    \includegraphics[width=0.5\textwidth]{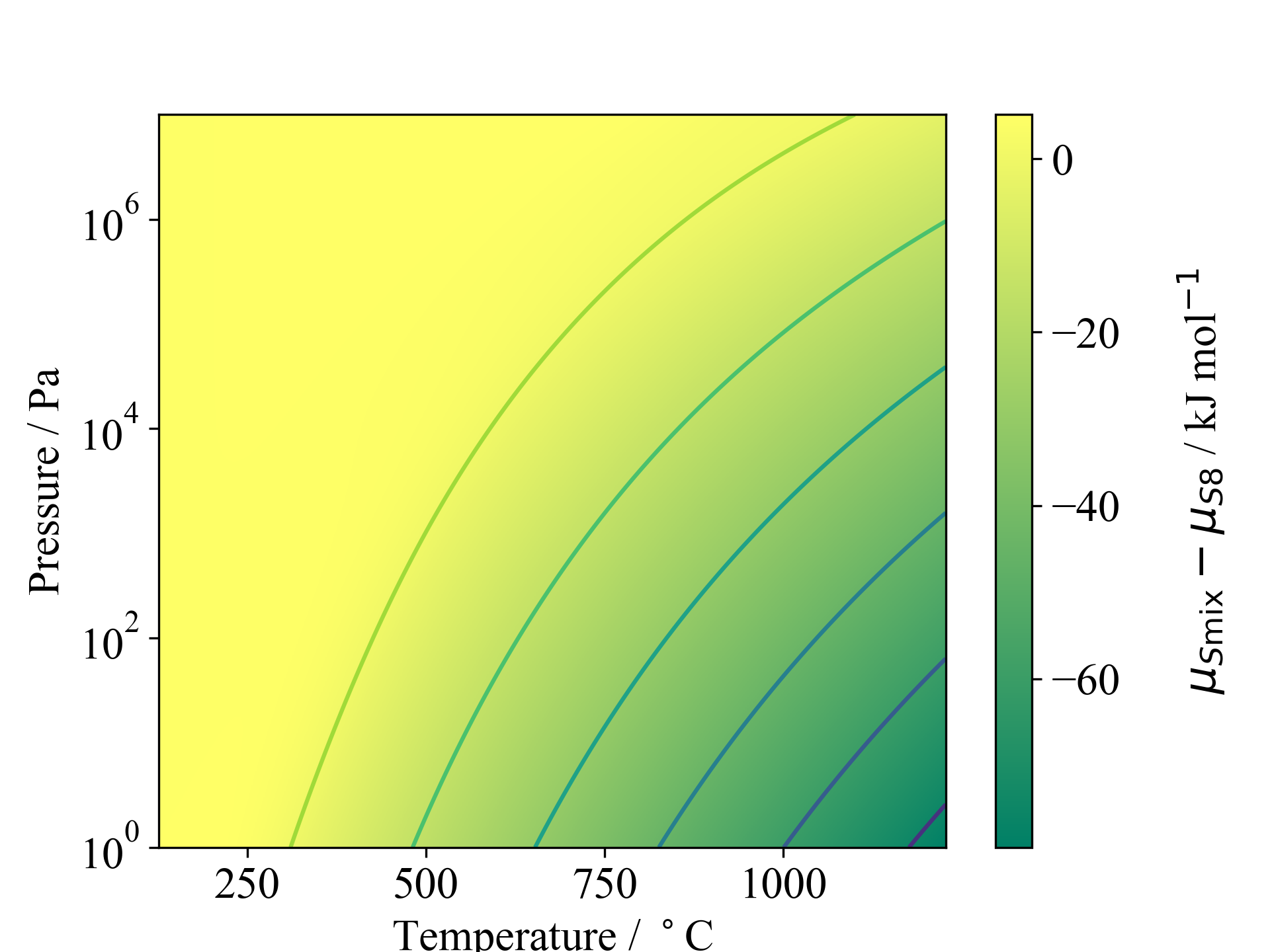}
    \caption{Comparison of the Gibbs free energy associated with a single allotrope sulfur vapour \ce{S8} and mixed allotrope $S_\mathrm{mix}$. The pressure is the partial pressure of the sulfur gas.}
    \label{fig:S8_mix}
\end{figure}

In Figure \ref{fig:S2_mix} we compare the predicted Gibbs free energies of a single allotrope \ce{S2} vapour with a mixed-allotrope vapour. In Figure \ref{fig:S8_mix} we compare the predicted Gibbs free energies of a single allotrope \ce{S8} vapour with a mixed-allotrope vapour. 
As outlined in the methods section of the main text the Gibbs free energy of the single-allotrope vapour is calculated using first-principles methods and tabulated experimental data, whilst the Gibbs free energy of the mixed-allotrope vapour is calculated using a previously published sulfur model\cite{jackson2016universal} parameterised with the hybrid PBE0 functional.\cite{carlo1999toward} 

As would be expected, in both Figure \ref{fig:S2_mix} and Figure \ref{fig:S8_mix} the models begin to diverge in the region where the published sulfur model predicts significant mixing of allotropes \ce{S2} to \ce{S8}.\cite{jackson2016universal} This coincides with the experimental conditions typically reported for \ce{BaZrS3} synthesis via annealing in sulfur vapour (temperatures between \SIrange{500}{600}{\celsius}, and pressures between \SIrange{1E3}{1E5}{\pascal}). This highlights the importance of moving beyond the approximation of a single sulfur species in models for modelling perovskite formation. The sulfur model is referenced to an \ce{S8} species so that there is closer agreement between \ce{S8} and S$
_\mathrm{mix}$ (in the low-T, high-P \ce{S8}-rich regions) than \ce{S2} and S$_\mathrm{mix}$ (in the high-T, low-P \ce{S2}-rich regions). Where the respective allotrope does not dominate, S$
_\mathrm{mix}$ is the more stable system (with a negative chemical potential of larger magnitude).


\newpage

\section{Sulfur transfer between binary precursors}

\begin{figure}[H]
    \centering
    \includegraphics[width=0.5\textwidth]{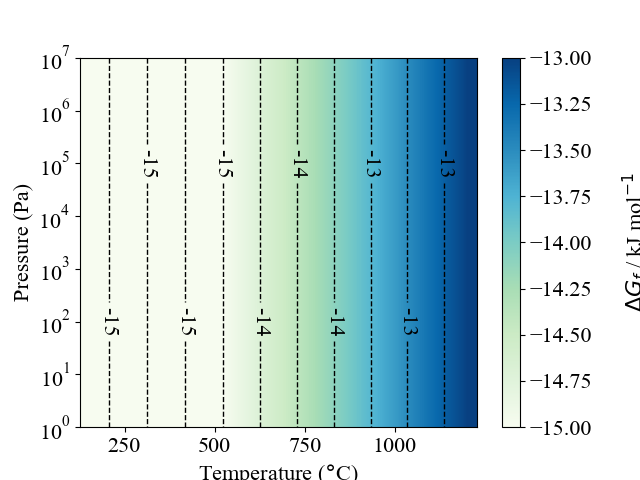}
    \caption{Gibbs free energy of formation for the reaction \ce{BaS3 + ZrS2 -> BaS2 + ZrS3}. As there is no sulfur-gas component, the pressure results from an inert gas or mechanical force.}
    \label{fig:BS3+ZS2_BS2+ZS3}
\end{figure}

Yang et al. report that after combining \ce{BaS3} and \ce{ZrS2} powders (with no sulfur excess) in a vacuum-sealed ampule a reaction is initiated to form \ce{BaS2} and \ce{ZrS3} at \SI{500}{\celsius}.\cite{yang2023low}
In Figure \ref{fig:BS3+ZS2_BS2+ZS3} we present the change in Gibbs free energy for the reaction \ce{BaS3 + ZrS2 -> BaS2 + ZrS3}. Across the whole temperature and pressure range the formation of \ce{BaS2} and \ce{ZrS3} is predicted to be thermodynamically favourable.

It is possible that sulfur gas is formed as an intermediate, so that the reaction is a two-step process:
\begin{chemequations}
\begin{equation}\label{eq:BS3+ZS2_BS2+ZS3_1}
    \ce{BaS3 + ZrS2 -> BaS2 + S(g) + ZrS2}
\end{equation}
\begin{equation}\label{eq:BS3+ZS2_BS2+ZS3_2}
    \ce{BaS2 + S(g) + ZrS2 -> BaS2 + ZrS3}
\end{equation}
\end{chemequations}

In Figure 4a of the main text we show that \ref{eq:BS3+ZS2_BS2+ZS3_1} is a forward reaction below a sulfur partial pressure of \SI{1}{\bar} at \SI{500}{\celsius}, which is realistic when there is no additional sulfur source.
Figure 4b shows that \ref{eq:BS3+ZS2_BS2+ZS3_2} is stabilised at sulfur partial pressures above \SI{1}{\milli\bar}.
Following formation of \ce{ZrS2} and \ce{BaS3}, there will be a kinetically limited reaction to form the final product \ce{BaZrS3}, as outlined in the main text.

We conclude that to stabilise the binary precursors \ce{BaS3} and \ce{ZrS2} at \SI{500}{\celsius} a sulfur partial pressure above \SI{1}{\bar} should be maintained so as to prevent \ref{eq:BS3+ZS2_BS2+ZS3_1}.

\newpage

\section{Relative stability of binary precursors in equilibrium with \ce{S2} or \ce{S8} vapours}

In the main text we present results for the binary Ba-S and Zr-S systems in equilibrium with a mixed-allotrope sulfur vapour. 
Here we compare the results for each system in equilibrium with either single-allotrope \ce{S2} or single-allotrope \ce{S8} vapours. 
A direct comparison between the mixed- and single-allotrope results is difficult as each simulation uses a different methodology. However we can compare single-allotrope \ce{S2} against single-allotrope \ce{S8} to better understand the range of behaviour that might be expected for systems that are out of thermodynamic equilibrium. 

\begin{figure}[H]
    \centering
    \includegraphics[width=0.6\textwidth]{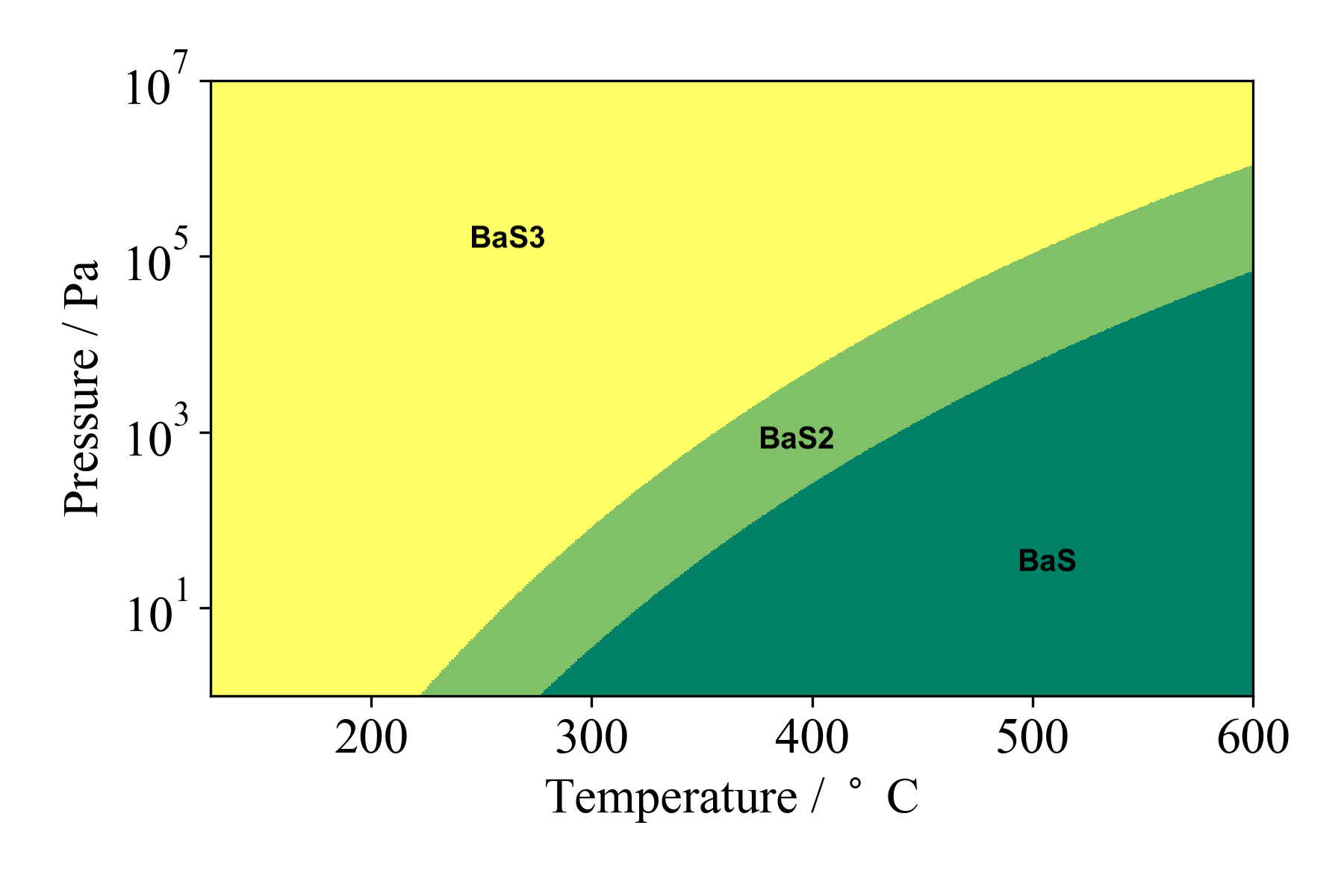}
    \caption{ Most stable material in the Ba-S system displayed as a function of temperature and pressure. The material is formed in equilibrium with a single allotrope \ce{S2}. Yellow corresponds to \ce{BaS3}, mid-green corresponds to \ce{BaS2} and dark green corresponds to \ce{BaS}. The pressure is the partial pressure of \ce{S2}.}
    \label{fig:Ba_S_stability_S2}
\end{figure}

\begin{figure}[H]
    \centering
    \includegraphics[width=0.6\textwidth]{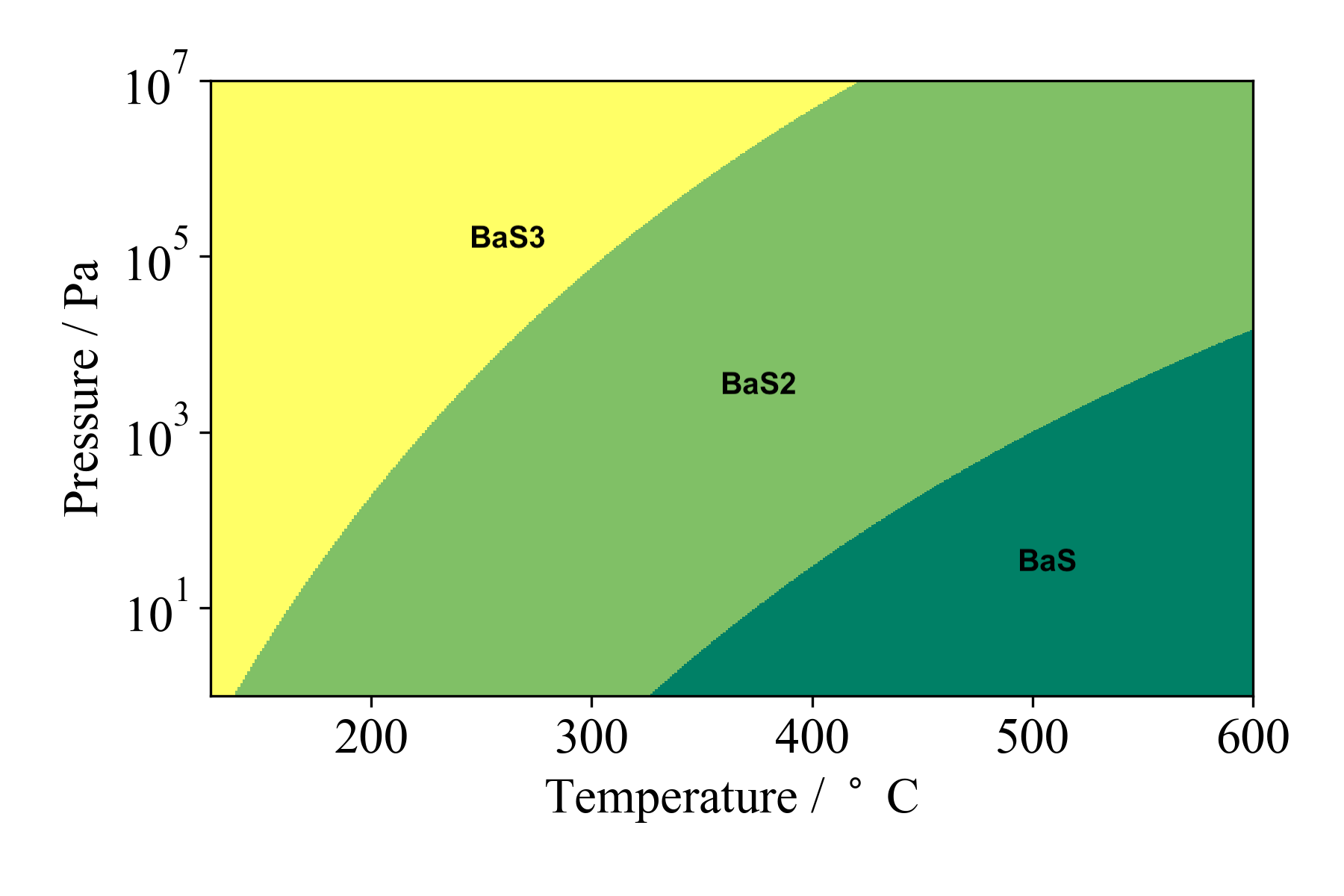}
    \caption{ Most stable material in the Ba-S system displayed as a function of temperature and pressure. The material is formed in equilibrium with a single allotrope \ce{S8}. Yellow corresponds to \ce{BaS3}, mid-green corresponds to \ce{BaS2} and dark green corresponds to \ce{BaS}. The pressure is the partial pressure of \ce{S8}.}
    \label{fig:Ba_S_stability_S8}
\end{figure}

\begin{figure}[H]
    \centering
    \includegraphics[width=0.6\textwidth]{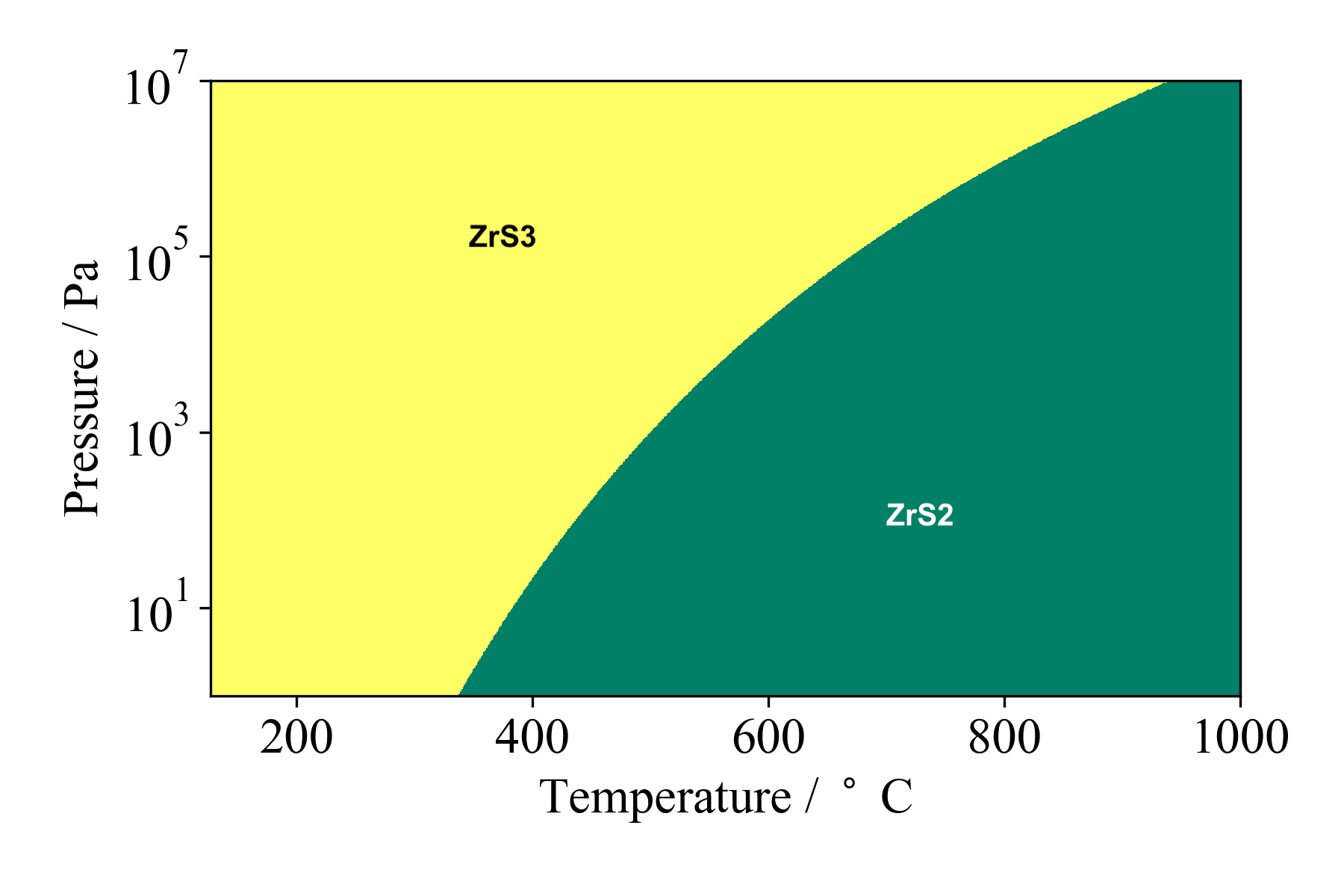}
    \caption{ Most stable material in the Zr-S system displayed as a function of temperature and pressure. The material is formed in equilibrium with a single allotrope \ce{S2}. Yellow corresponds to \ce{ZrS3} and dark-green corresponds to \ce{ZrS2}. The pressure is the partial pressure of \ce{S2}.}
    \label{fig:Zr_S_stability_S2}
\end{figure}

\begin{figure}[H]
    \centering
    \includegraphics[width=0.6\textwidth]{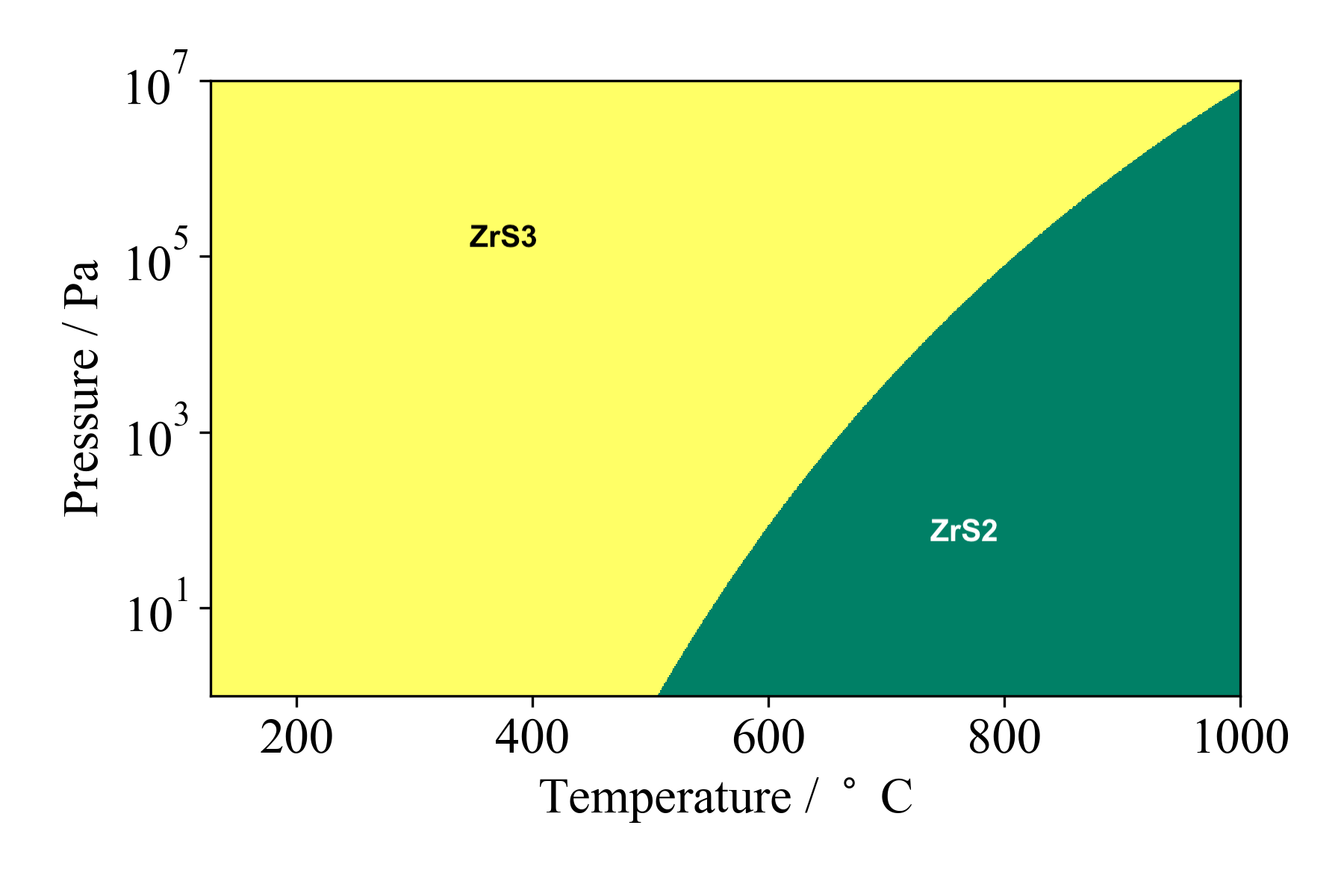}
    \caption{Most stable material in the Zr-S system displayed as a function of temperature and pressure. The material is formed in equilibrium with a single allotrope \ce{S8}. Yellow corresponds to \ce{ZrS3} and dark-green corresponds to \ce{ZrS2}. The pressure is the partial pressure of \ce{S8}.}
    \label{fig:Zr_S_stability_S8}
\end{figure}

We find that the position of the co-existance curves (where two systems have equal Gibbs free energy) are sensitive to sulfur vapour type, with \ce{S2} vapour promoting the formation of \ce{BaS3} across an increased temperature and pressure range compared to \ce{S8} (Figures \ref{fig:Ba_S_stability_S2} and \ref{fig:Ba_S_stability_S8}). 
This suggests that a high-temperature sulfur source producing \ce{S2} may promote the formation of sulfur-rich species if it can react with the metal precursor before equilibrating to a more stable, mixed-allotrope vapour.
For the Zr-S system we find that \ce{S2} vapour hinders the formation of sulfur rich \ce{ZrS3}. For example, \ce{ZrS3} is thermodynamically favourable across all partial pressures at \SI{500}{\celsius} in \ce{S8}, whilst this is reduced to \SI{350}{\celsius} in \ce{S2} (Figures \ref{fig:Zr_S_stability_S2} \ref{fig:Zr_S_stability_S8}). 

\newpage
\section{Extension of model to other reaction processes}

There are a large number of mass-balanced reactions which are possible for the Ba-Zr-S system we are modelling. 
We do not present the predicted Gibbs free energy of formation for all processes, but have highlighted those of particular interest and importance. 
A more complete set of reactions are outlined in the Jupyter Notebook associated with this paper,\cite{PaperRepo}
and we provide the first-principles data,\cite{PaperRepo, NOMAD} and software\cite{ThermoPot} required to consider any reaction for the set of Ba-Zr-S materials in this study.

We also note that it is not necessary to consider all possible product and reactant combinations; for example if we know that the reactions \ce{A + B -> C + D} and \ce{C + D -> E} are favourable (has a negative Gibbs free energy), we can deduce that \ce{A + B -> E} will also be favourable. This reduces the number of reactions to a more manageable size.

\begin{figure}[H]
    \centering
    \includegraphics[width=0.5\textwidth]{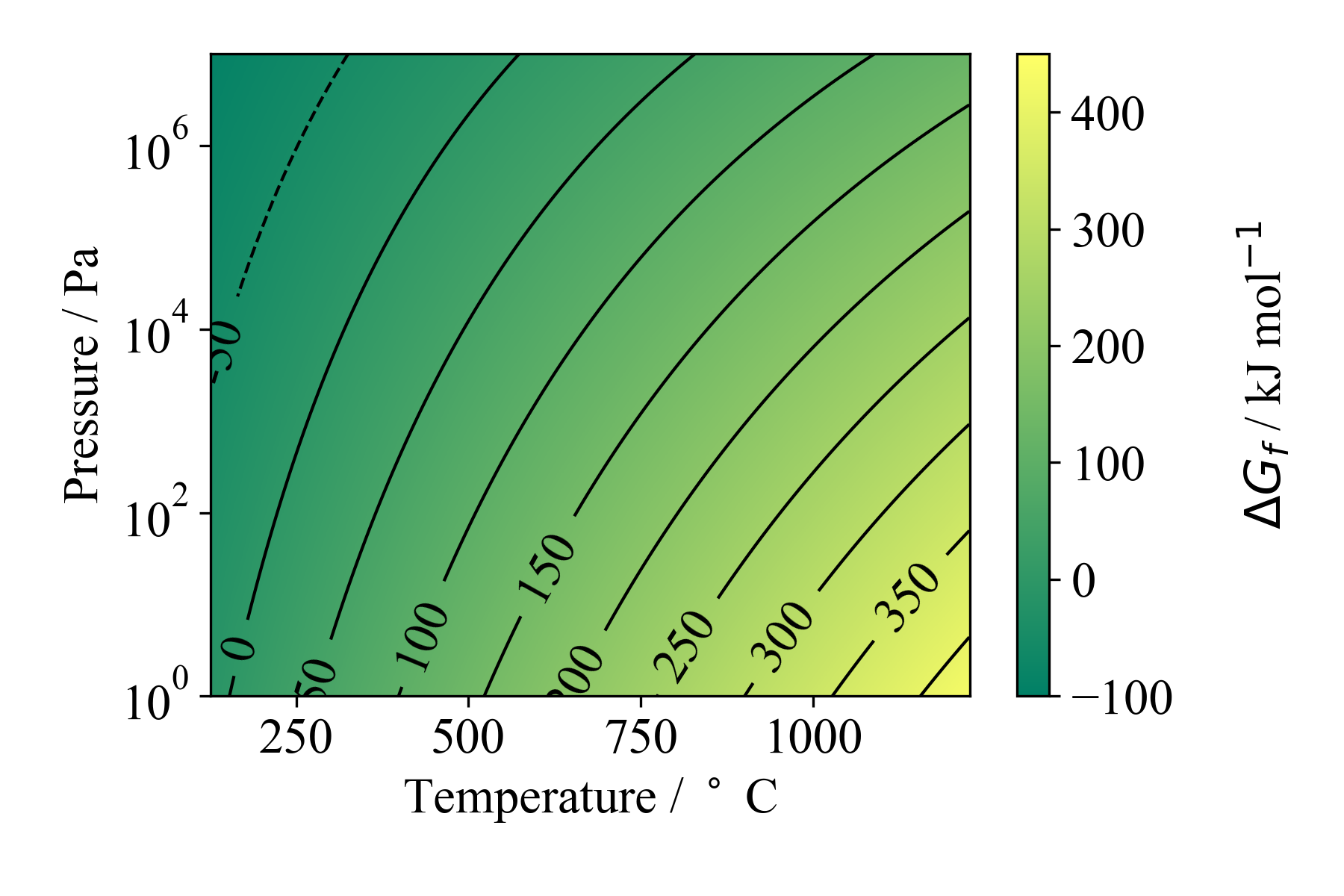}
    \caption{Gibbs free energy of the reaction $\frac{3}{2}$S$_2$ + BaZrS$_3$ $\rightarrow$ BaS$_3$ + ZrS$_3$ }
    \label{fig:extension_1}
\end{figure}

In Figure \ref{fig:extension_1} we show one result highlighted in the main text: formation of \ce{BaZrS3} in an atmosphere with the single-allotrope vapour \ce{S2}, where we predict degradation at lower temperatures and higher partial pressures. For example, Figure \ref{fig:extension_1} shows a thermodynamic driving force towards degradation at \SI{1}{\bar} and \SI{400}{\celsius}. However this region lies above the saturation vapour pressure for sulfur, where the ideal gas model for sulfur vapour is no longer valid.

\newpage
\section{Comparison of exchange-correlation functionals}

\begin{figure}[H]
    \centering
    \includegraphics[width=0.5\textwidth]{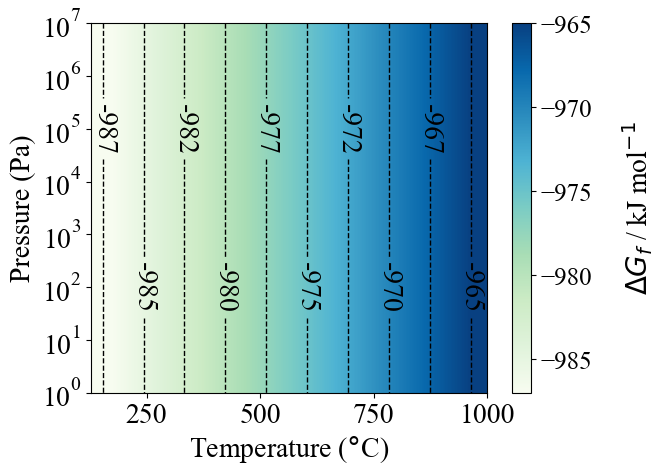}
    \caption{Gibbs free energy of Ba + Zr + 3S(s) $\rightarrow$ BaZrS$_3$, calculated using the hybrid HSE06 exchange-correlation functional.}
    \label{fig:C1}
\end{figure}

\begin{figure}[H]
    \centering
    \includegraphics[width=0.5\textwidth]{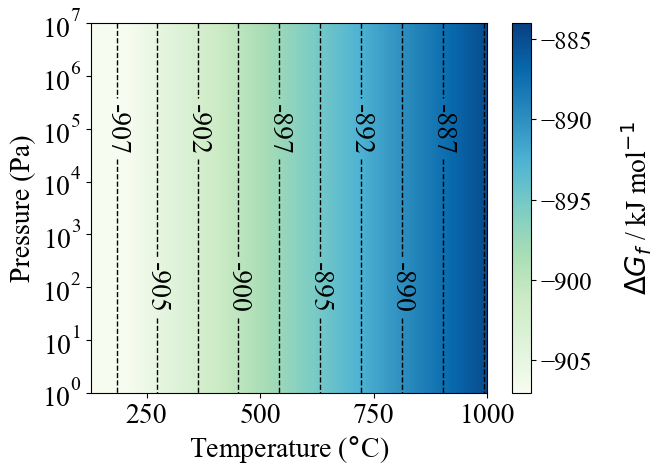}
    \caption{Gibbs free energy of Ba + Zr + 3S(s) $\rightarrow$ BaZrS$_3$, calculated using the PBEsol exchange-correlation functional.}
    \label{fig:C2}
\end{figure}

\begin{figure}[H]
    \centering
    \includegraphics[width=0.5\textwidth]{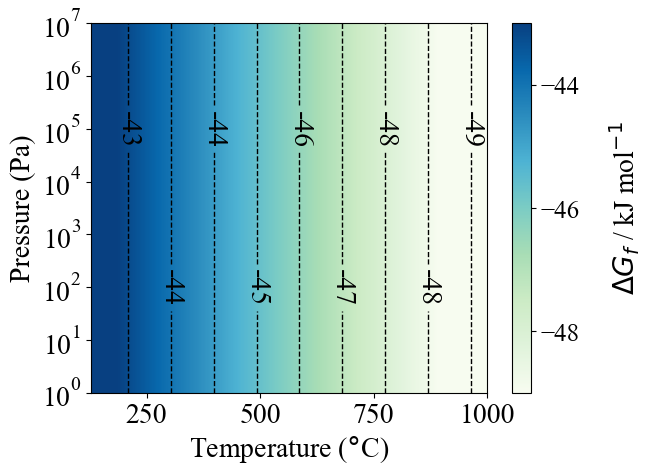}
    \caption{Gibbs free energy of BaS + \ce{ZrS2} $\rightarrow$ \ce{BaZrS3}, calculated using the hybrid HSE06 exchange-correlation functional.}
    \label{fig:C3}
\end{figure}

\begin{figure}[H]
    \centering
    \includegraphics[width=0.5\textwidth]{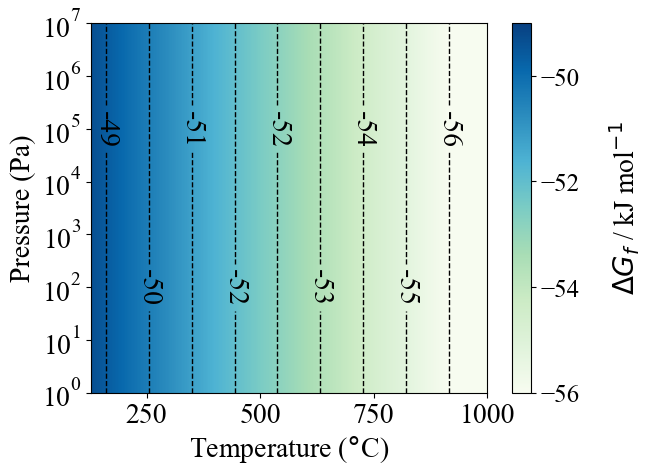}
    \caption{Gibbs free energy of BaS + \ce{ZrS2} $\rightarrow$ \ce{BaZrS3}, calculated using the PBEsol exchange-correlation functional.}
    \label{fig:C4}
\end{figure}

\begin{figure}[H]
    \centering
    \includegraphics[width=0.5\textwidth]{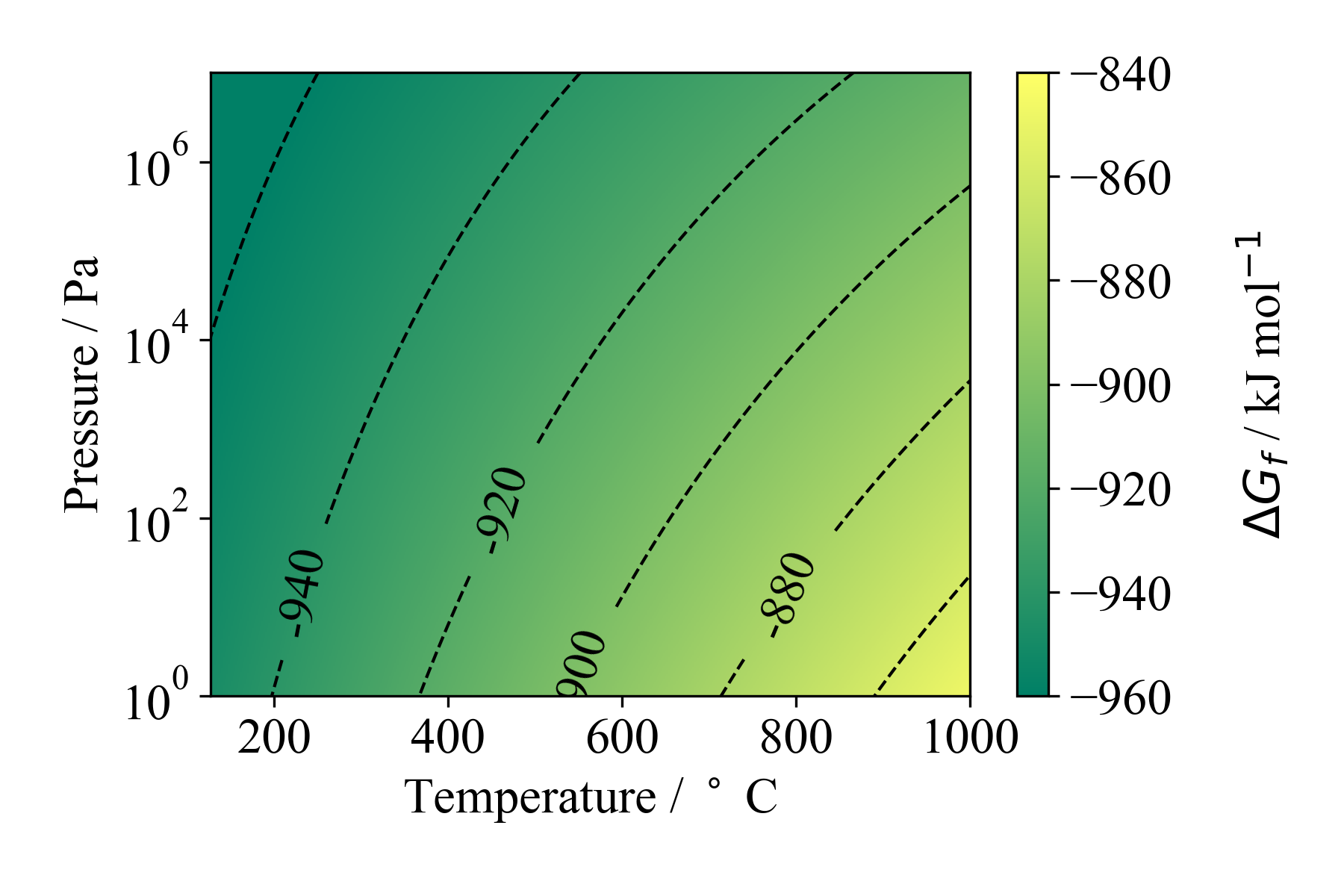}
    \caption{Gibbs free energy of Ba + Zr + $\frac{3}{8}$S$_8\mathrm{(g)} \rightarrow \ce{BaZrS3}$, calculated using the hybrid HSE06 exchange-correlation functional.}
    \label{fig:C5}
\end{figure}

\begin{figure}[H]
    \centering
    \includegraphics[width=0.5\textwidth]{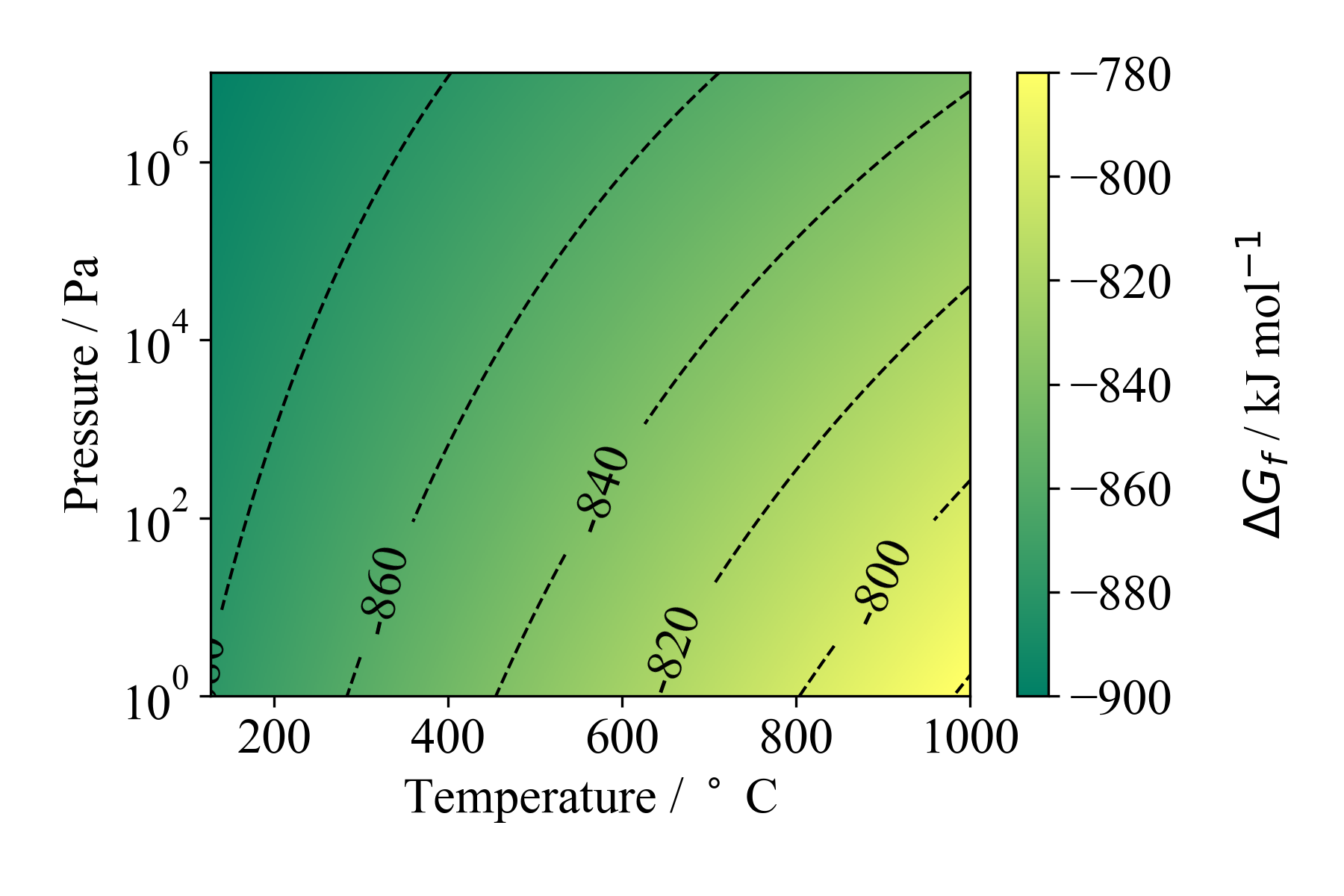}
    \caption{Gibbs free energy of Ba + Zr + $\frac{3}{8}$S$_8\mathrm{(g)} \rightarrow \ce{BaZrS_3}$, calculated using the PBEsol exchange-correlation functional.}
    \label{fig:C6}
\end{figure}

\begin{figure}[H]
    \centering
    \includegraphics[width=0.5\textwidth]{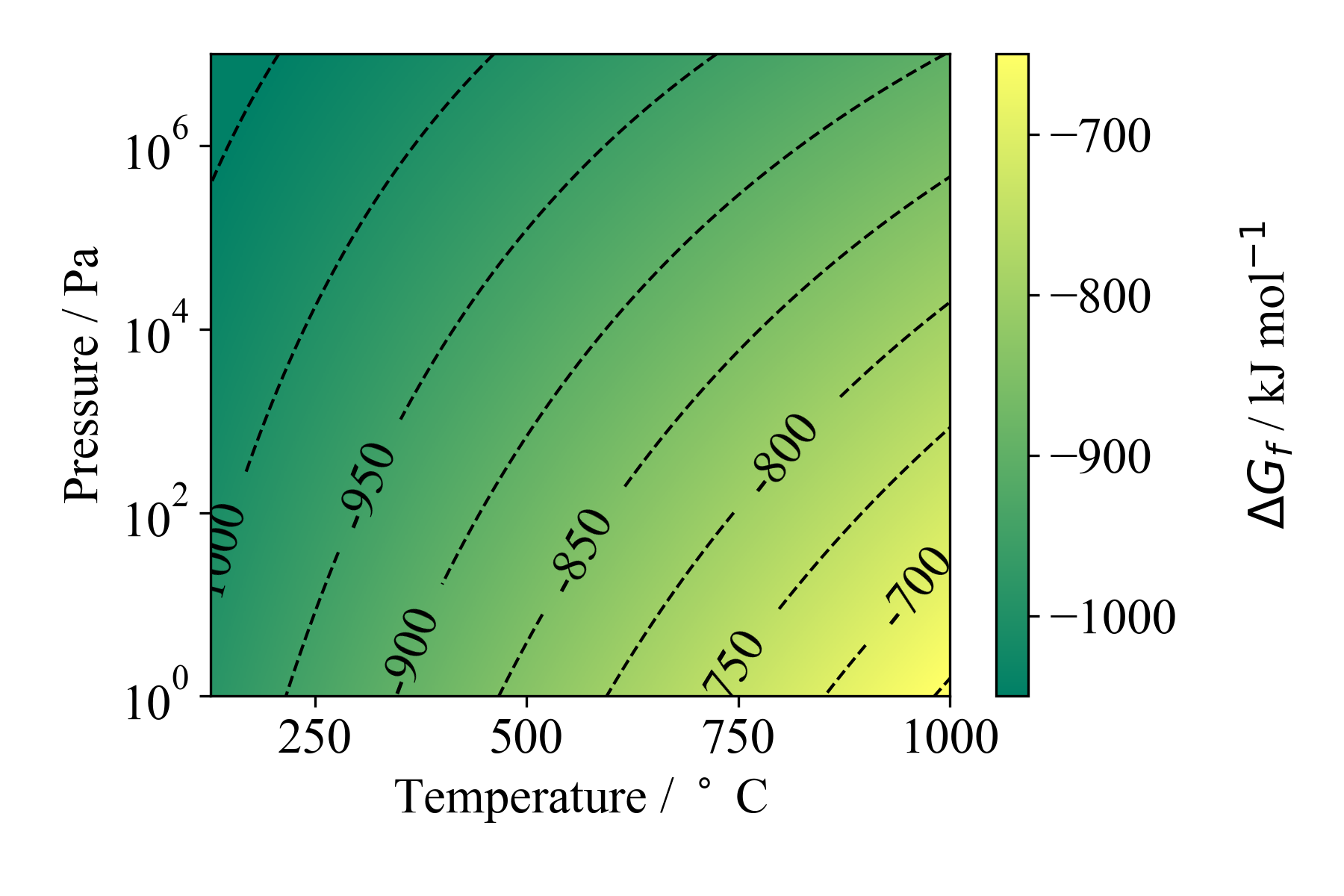}
    \caption{Gibbs free energy of Ba + Zr + $\frac{3}{2}$S$_2\mathrm{(g)} \rightarrow \ce{BaZrS_3}$, calculated using the hybrid HSE06 exchange-correlation functional.}
    \label{fig:C7}
\end{figure}

\begin{figure}[H]
    \centering
    \includegraphics[width=0.5\textwidth]{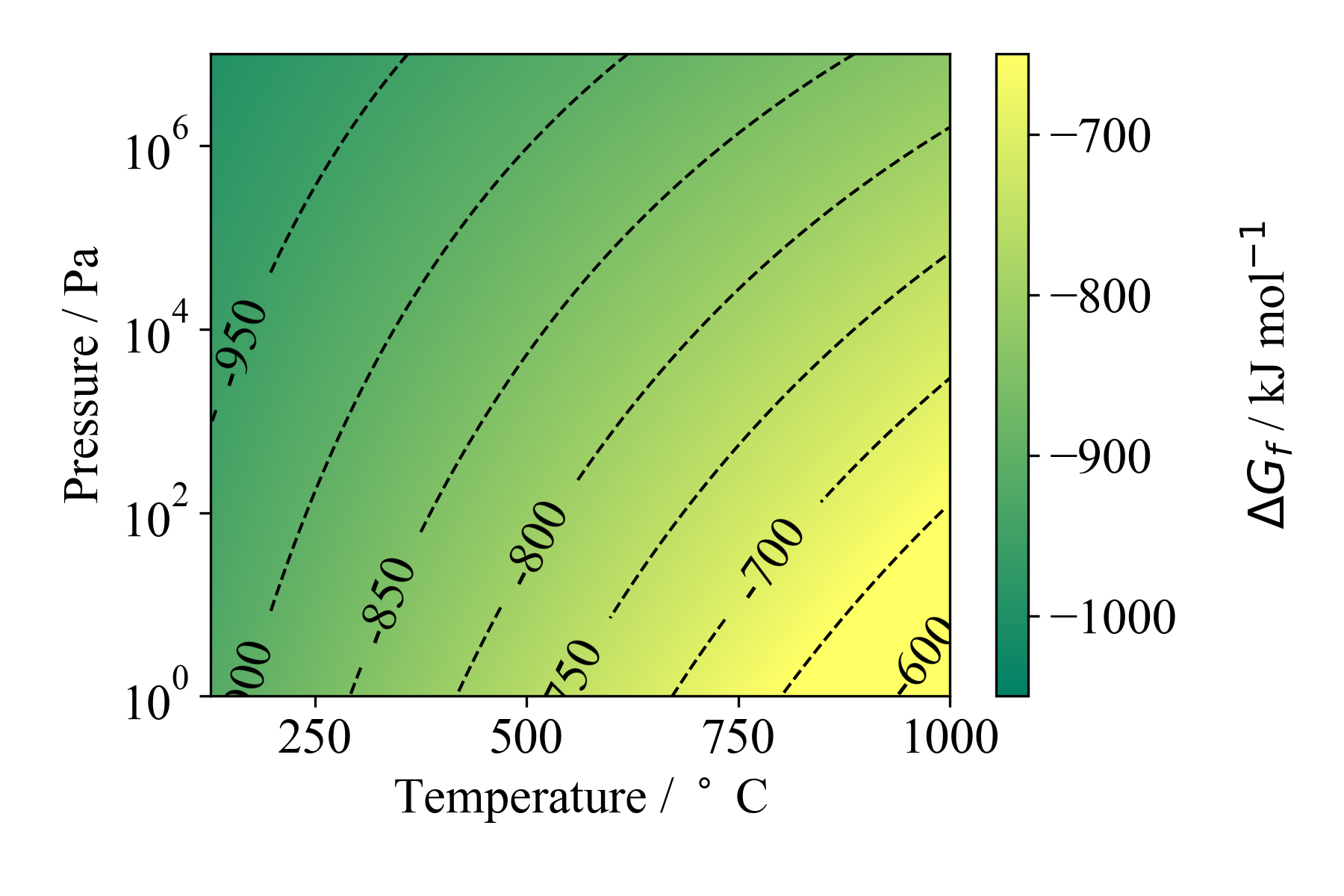}
    \caption{Gibbs free energy of Ba + Zr + $\frac{3}{2}$S$_2\mathrm{(g)} \rightarrow \ce{BaZrS_3}$ , calculated using the PBEsol exchange-correlation functional.}
    \label{fig:C8}
\end{figure}

\begin{figure}[H]
    \centering
    \includegraphics[width=0.5\textwidth]{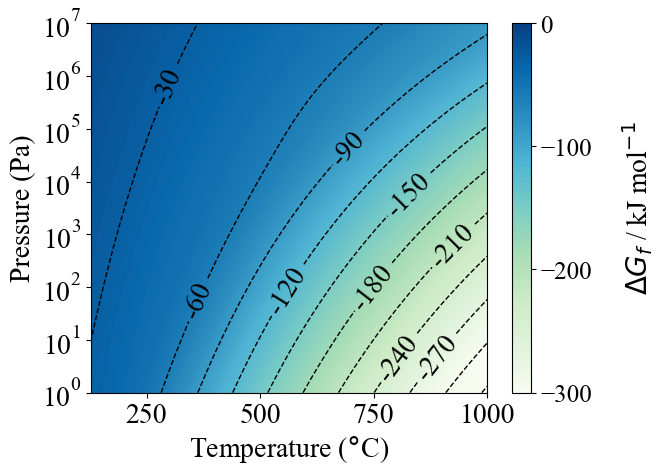}
    \caption{Gibbs free energy of \ce{BaZrS3} + 3\ce{S$_\mathrm{mix}$}\textrm{(g)} $\rightarrow$ \ce{ZrS3} + \ce{BaS3}, calculated using the hybrid HSE06 exchange-correlation functional.}
    \label{fig:C9}
\end{figure}

\begin{figure}[H]
    \centering
    \includegraphics[width=0.5\textwidth]{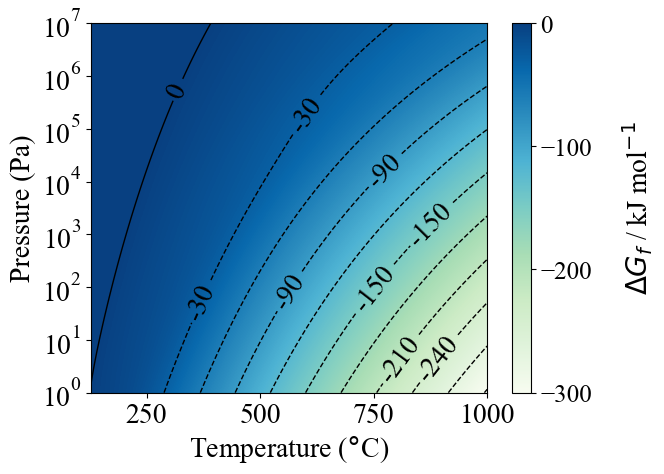}
    \caption{Gibbs free energy of \ce{BaZrS3} + 3\ce{S$_\mathrm{mix}$}\textrm{(g)} $\rightarrow$ \ce{ZrS3} + \ce{BaS3}, calculated using the PBEsol exchange-correlation functional.}
    \label{fig:C10}
\end{figure}

\begin{figure}[H]
    \centering
    \includegraphics[width=0.5\textwidth]{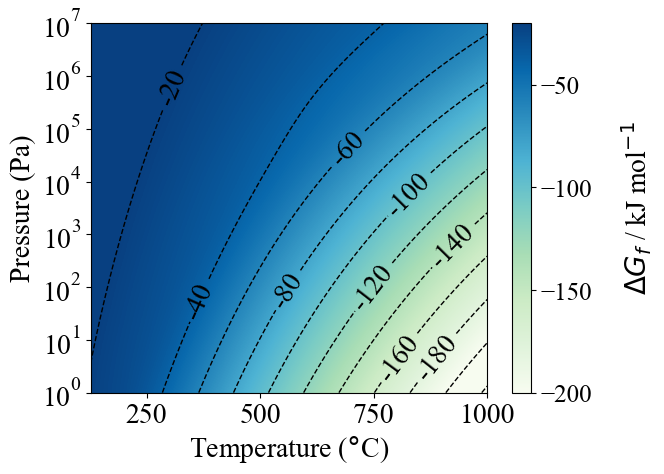}
    \caption{\ce{BaZrS3} + 2\ce{S$_\mathrm{mix}$}\textrm{(g)} $\rightarrow$ \ce{ZrS3} + \ce{BaS2}, calculated using the hybrid HSE06 exchange-correlation functional.}
    \label{fig:C11}
\end{figure}

\begin{figure}[H]
    \centering
    \includegraphics[width=0.5\textwidth]{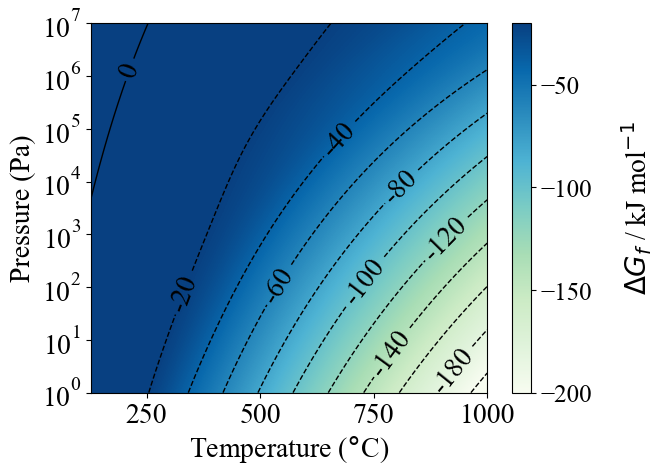}
    \caption{\ce{BaZrS3} + 2\ce{S$_\mathrm{mix}$}\textrm{(g)} $\rightarrow$ \ce{ZrS3} + \ce{BaS2}, calculated using the PBEsol exchange-correlation functional.}
    \label{fig:C12}
\end{figure}

\begin{figure}[H]
    \centering
    \includegraphics[width=0.5\textwidth]{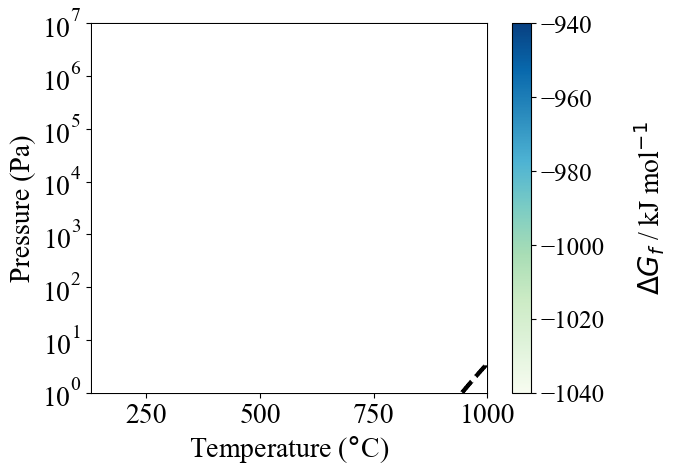}
    \caption{The coexistance curve for single allotrope \ce{S2} vapour
and single allotrope \ce{S8} vapour, with total energies calculated using the hybrid HSE06 exchange-correlation functional. The dashed line indicates where the
chemical potentials of the sulfur gas allotropes on a per-atom basis are equal, $\mu_{S8} = \mu_{S2}$. \ce{S8} dominates in the top left region, {S2} dominates in the bottom right.}
    \label{fig:C13}
\end{figure}

\begin{figure}[H]
    \centering
    \includegraphics[width=0.5\textwidth]{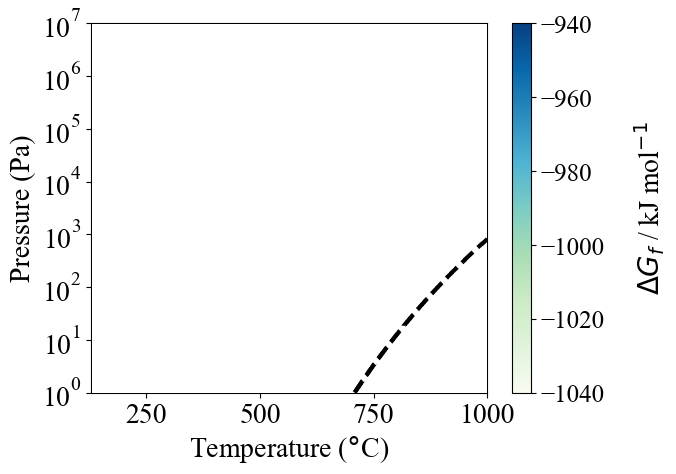}
    \caption{The coexistance curve for single allotrope \ce{S2} vapour
and single allotrope \ce{S8} vapour, with total energies calculated using the PBEsol exchange-correlation functional. The dashed line indicates where the
chemical potentials of the sulfur gas allotropes on a per-atom basis are equal, $\mu_{S8} = \mu_{S2}$. \ce{S8} dominates in the top left region, {S2} dominates in the bottom right.}
    \label{fig:C14}
\end{figure}

Figure \ref{fig:C1} to \ref{fig:C14} we show results for the Gibbs free energy of formations calculated using the exchange-correlation functionals hybrid HSE06\cite{krukau2006influence} and PBEsol\cite{perdew2008restoring} (whilst SCAN\cite{sun2015strongly} is used in the main text). 

In Figures \ref{fig:C1} to \ref{fig:C12} we see that the perovskite is predicted to remain stable with respect to elemental and binary materials, irrespective of the exchange-correlation functional used.
In Figures \ref{fig:C13} and \ref{fig:C14} We show that the \ce{S2}/\ce{S8} co-existance curve (where the
chemical potentials of the sulfur gas allotropes on a per-atom basis are equal, $\mu_{S_8} = \mu_{S_2}$) is highly sensitive to the functional used. Compared to the PBEsol functional, the hybrid HSE06 functional predicts that S2 will dominate over a smaller region at high-temperature and low-pressure. Both disagree with a previously published higher-accuracy result using the PBE0 functional,\cite{jackson2016universal} which shows \ce{S2} will dominate in the temperature and pressure conditions typically used for \ce{BaZrS3} synthesis (further discussion on this is provided in the Methods section of the main text).

All of the reactions considered in the main text can be analysed using the HSE06 or PBEsol functionals using the dataset\cite{NOMAD,PaperRepo}, workflows\cite{PaperRepo} and software\cite{ThermoPot} accompanying this paper.


\newpage 

\section{Electronic bandstructures}

\begin{figure}[H]
    \centering
    \includegraphics[width=\textwidth]{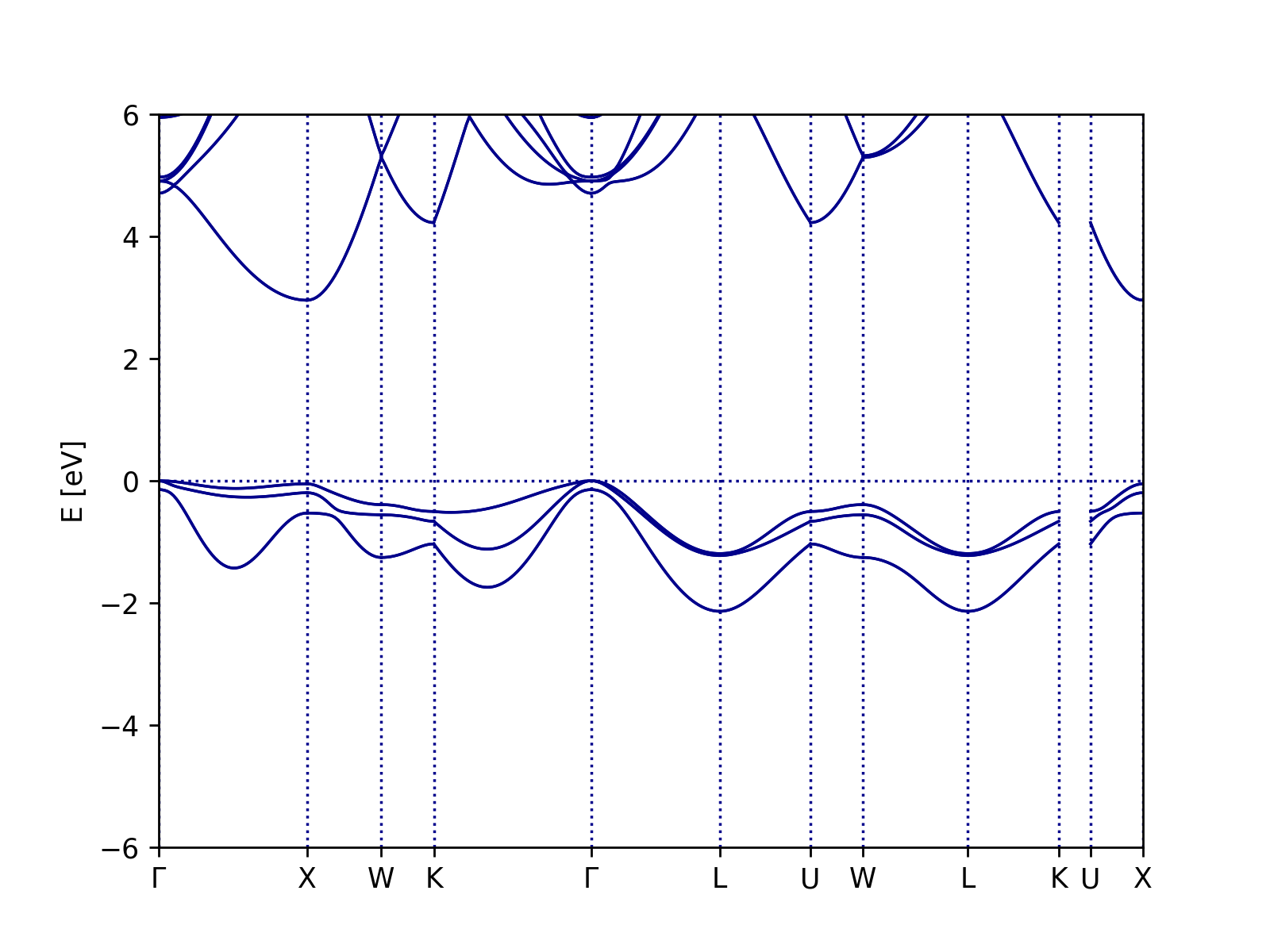}
    \caption{BaS ($Fm\Bar{3}m$) electronic band structure}
    \label{fig:E1}
\end{figure}

\begin{figure}[H]
    \centering
    \includegraphics[width=\textwidth]{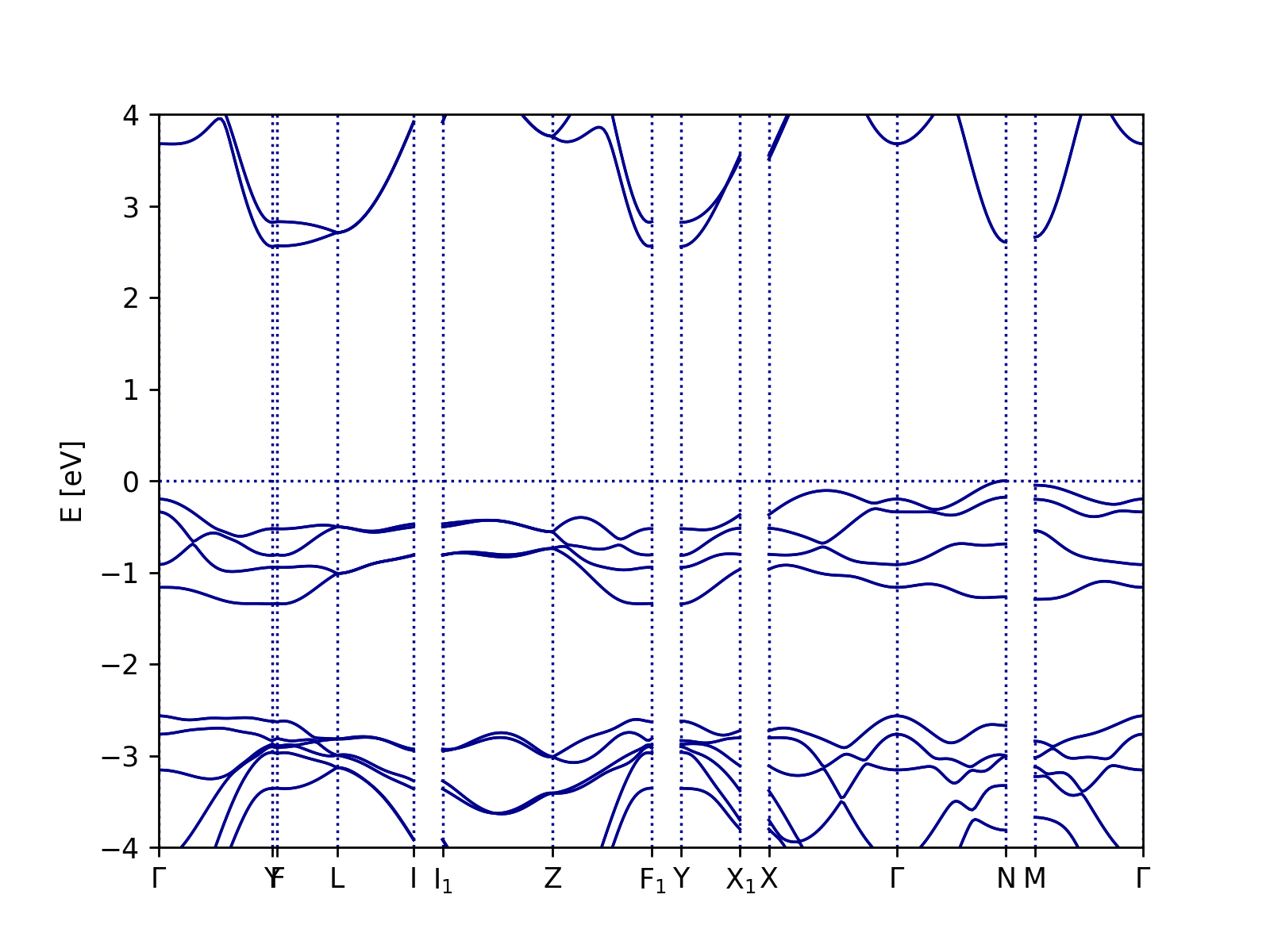}
    \caption{\ce{BaS2} (C2/c) electronic band structure}
    \label{fig:E2}
\end{figure}

\begin{figure}[H]
    \centering
    \includegraphics[width=\textwidth]{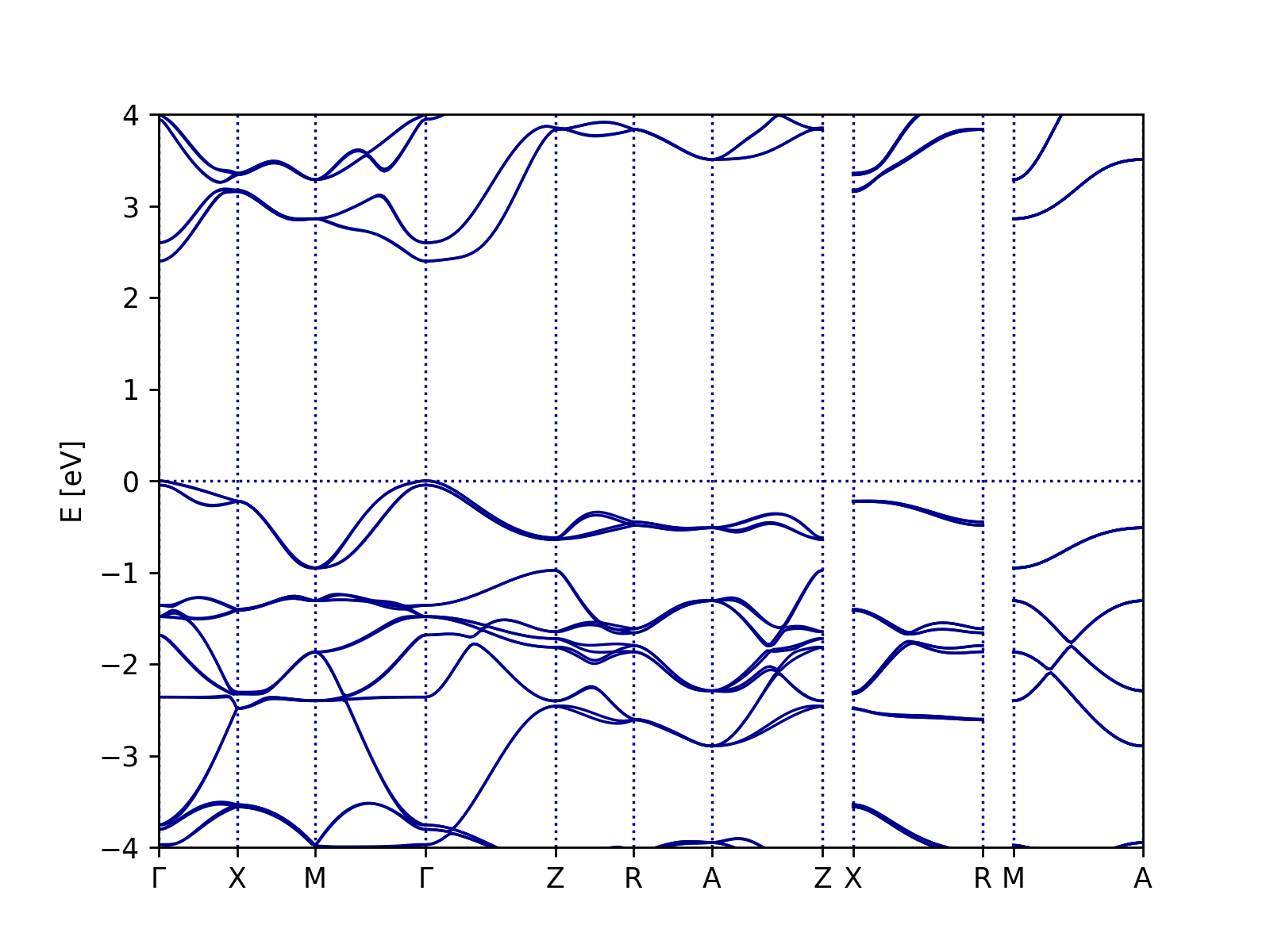}
    \caption{\ce{BaS3} ($P\Bar{4}2_1m$) electronic band structure}
    \label{fig:E3}
\end{figure}

\begin{figure}[H]
    \centering
    \includegraphics[width=\textwidth]{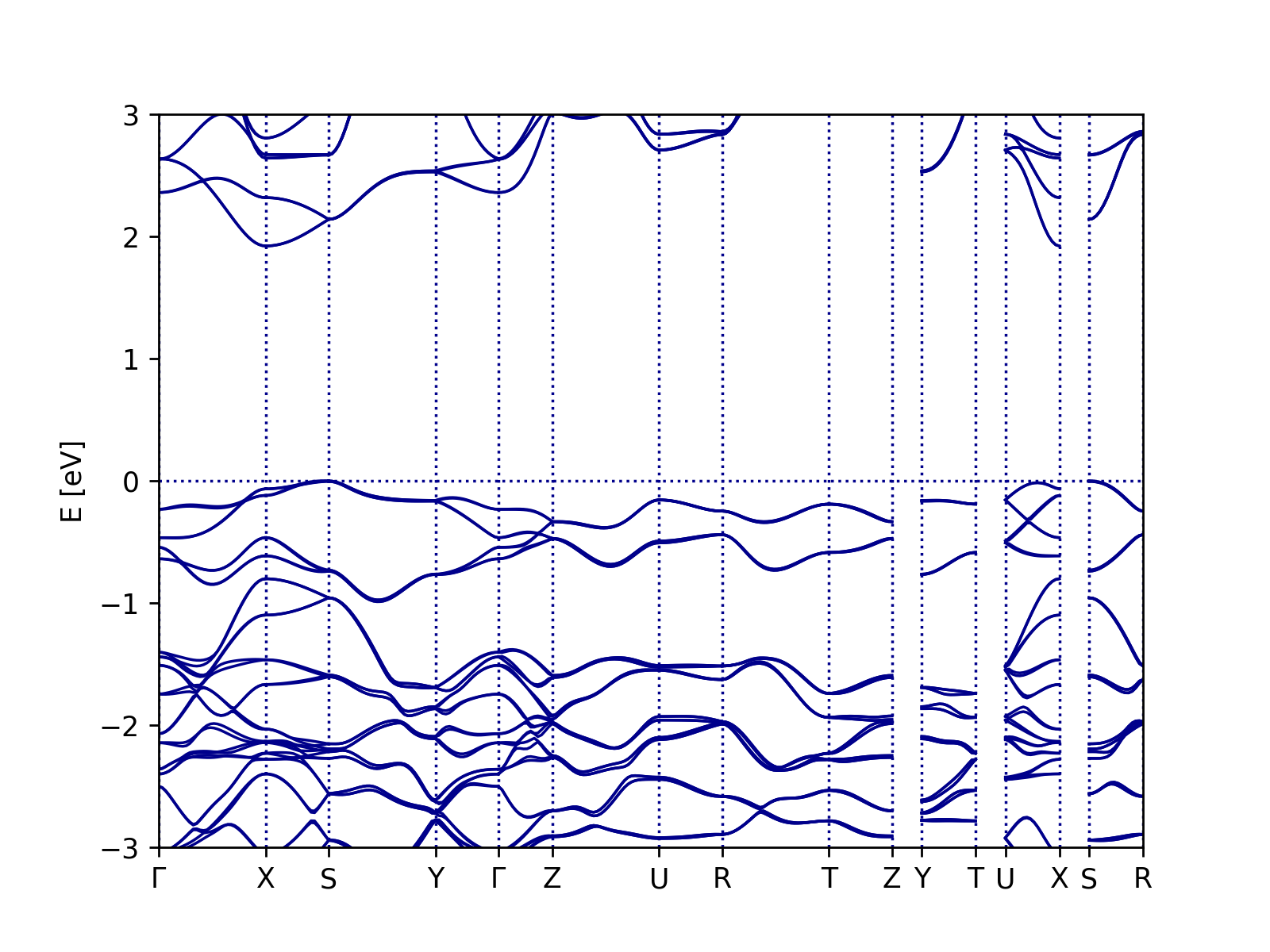}
    \caption{\ce{BaS3} ($P2_12_12$) electronic band structure}
    \label{fig:E4}
\end{figure}

\begin{figure}[H]
    \centering
    \includegraphics[width=\textwidth]{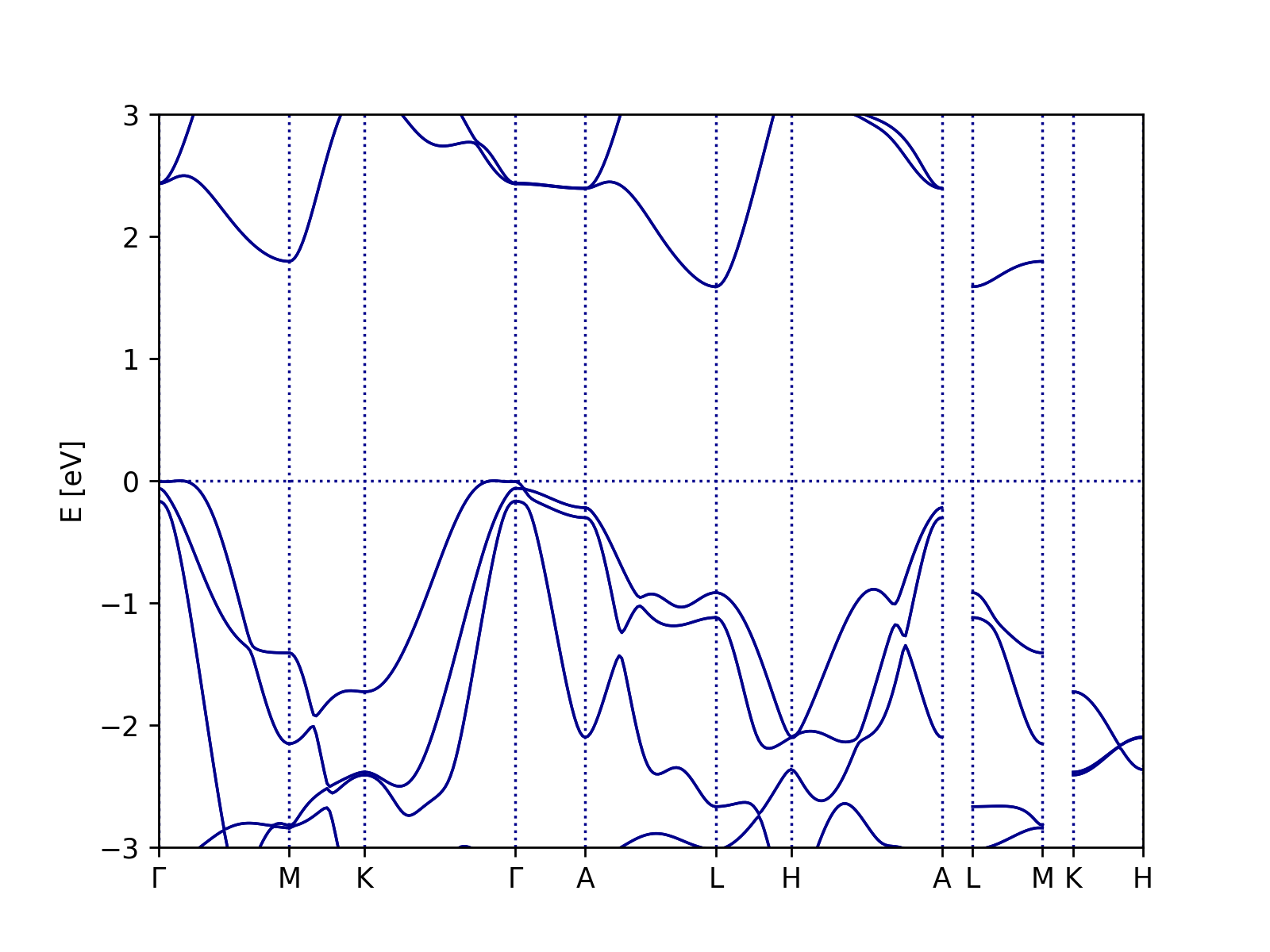}
    \caption{\ce{ZrS2} ($P\Bar{3}$m1) electronic band structure}
    \label{fig:E7}
\end{figure}

\begin{figure}[H]
    \centering
    \includegraphics[width=\textwidth]{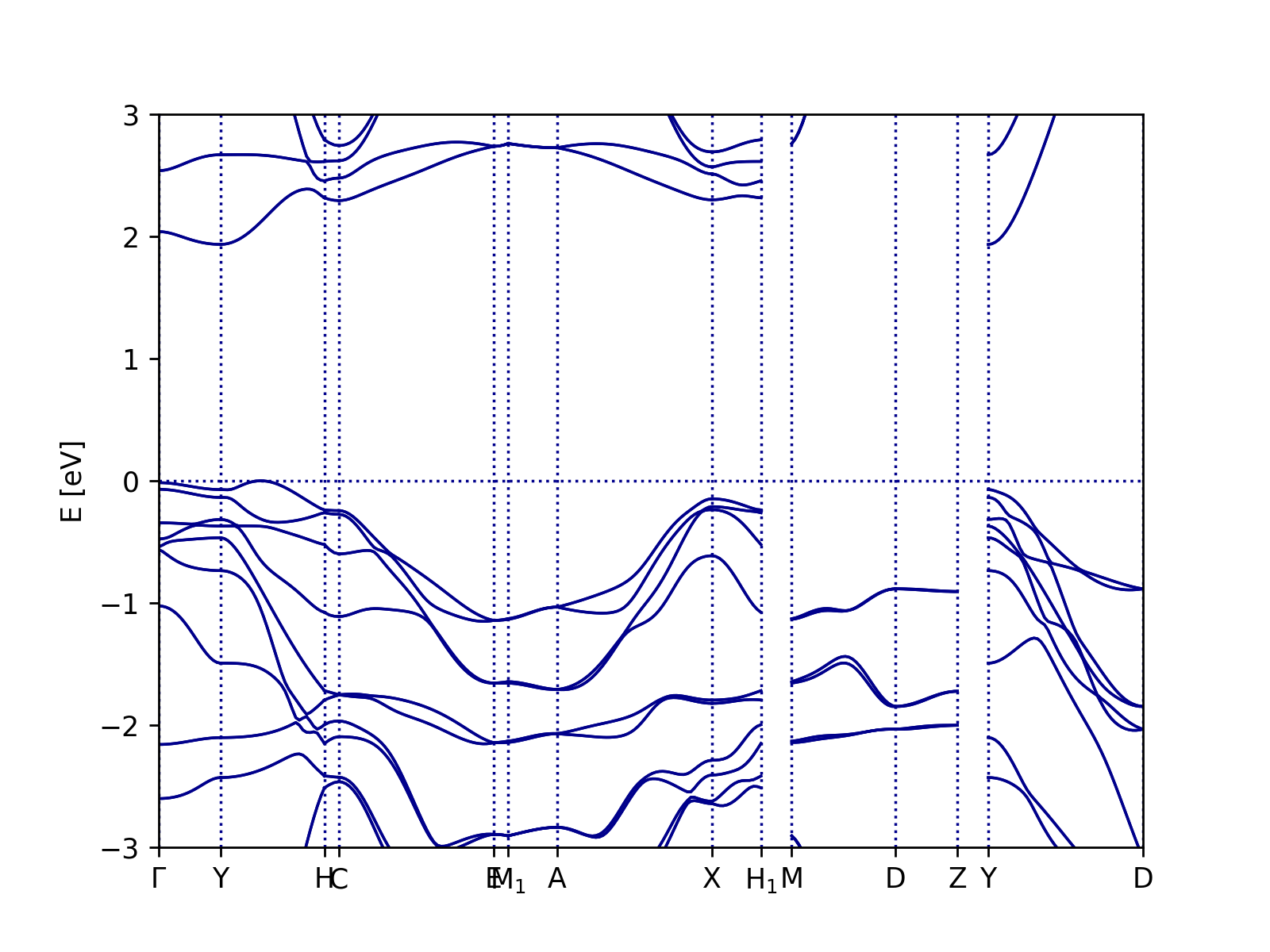}
    \caption{\ce{ZrS3} ($P2_1/m$) electronic band structure}
    \label{fig:E8}
\end{figure}

\begin{figure}[H]
    \centering
    \includegraphics[width=\textwidth]{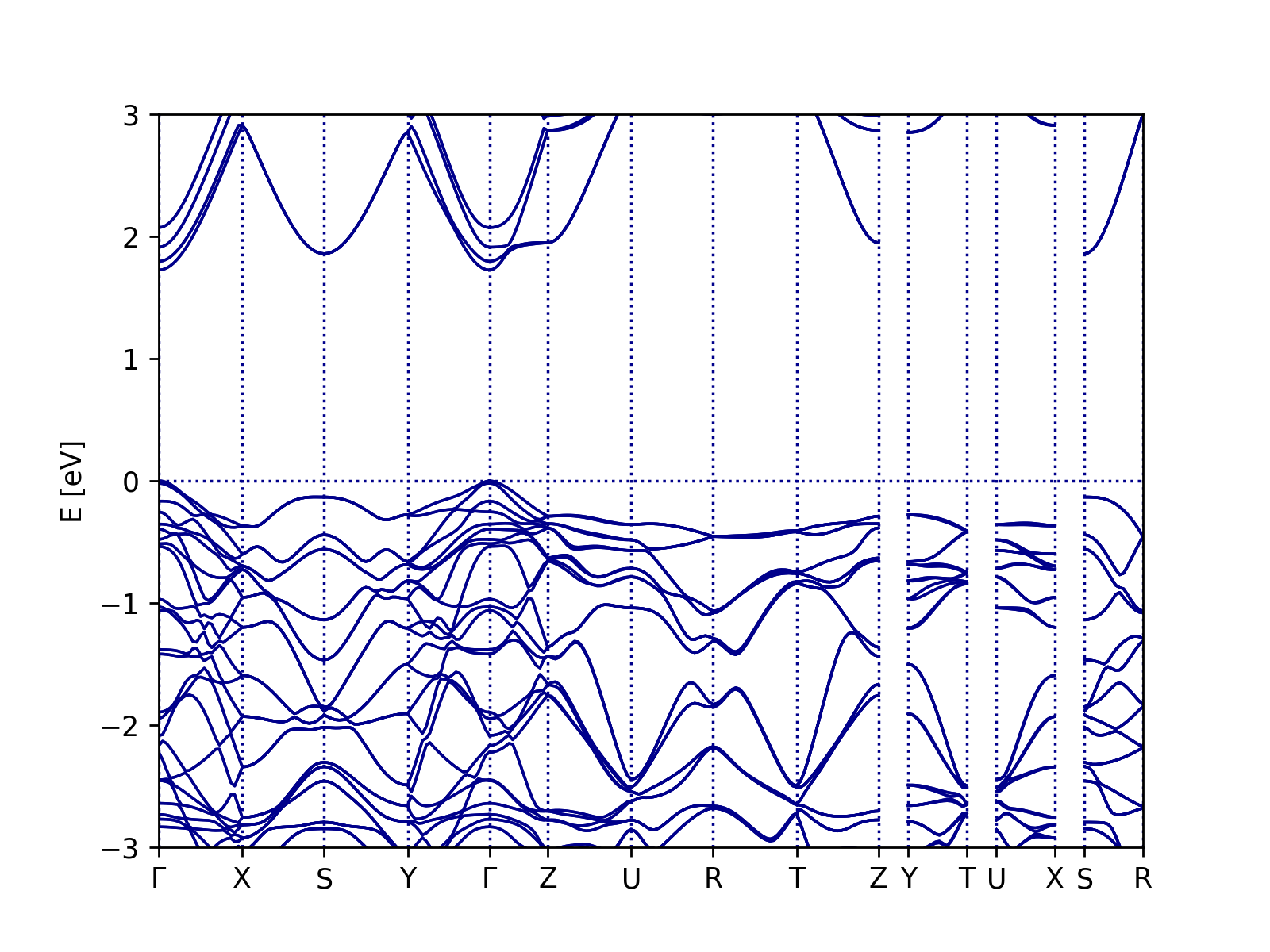}
    \caption{\ce{BaZrS3} ($Pnma$) electronic band structure}
    \label{fig:E9}
\end{figure}

Figures \ref{fig:E1} to \ref{fig:E9} show the electronic bandstructures for \ce{BaZrS3} and the binary materials selected in this study, calculated using the HSE06 functional. The bandstructures are generated from density functional theory calculations, as outlined in the methods section of the main text. They have been plotted using \texttt{aimstools} (github.com/romankempt/aimstools).


\newpage

\section{Phonon bandstructures}

\begin{figure}[H]
    \centering
    \includegraphics[width=\textwidth]{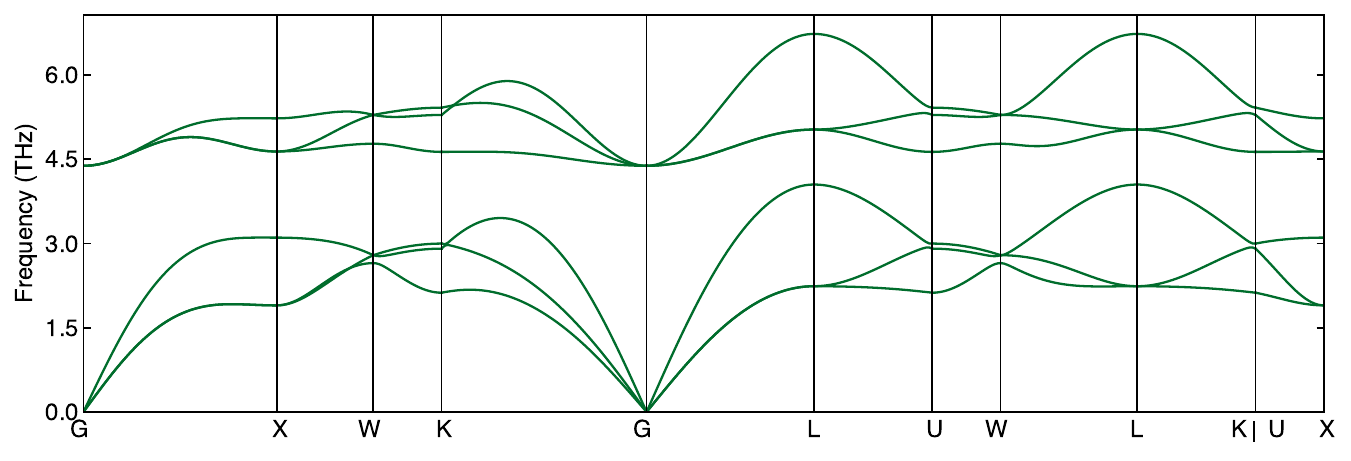}
    \caption{\ce{BaS} ($Fm\Bar{3}m$) phonon band structure}
    \label{fig:P1}
\end{figure}

\begin{figure}[H]
    \centering
    \includegraphics[width=\textwidth]{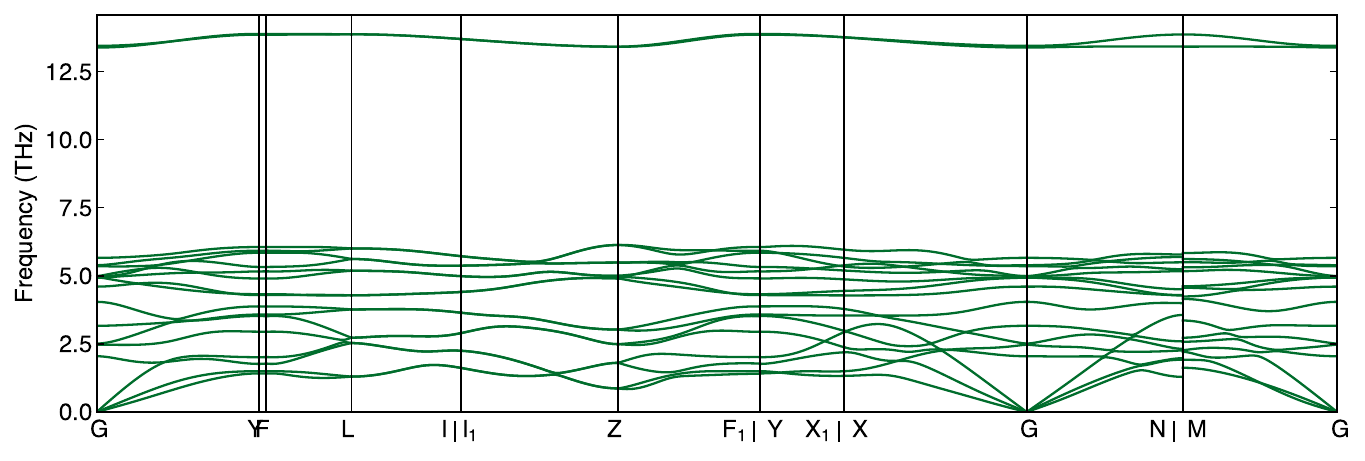}
    \caption{\ce{BaS2} ($C2/c$) phonon band structure}
    \label{fig:P2}
\end{figure}

\begin{figure}[H]
    \centering
    \includegraphics[width=\textwidth]{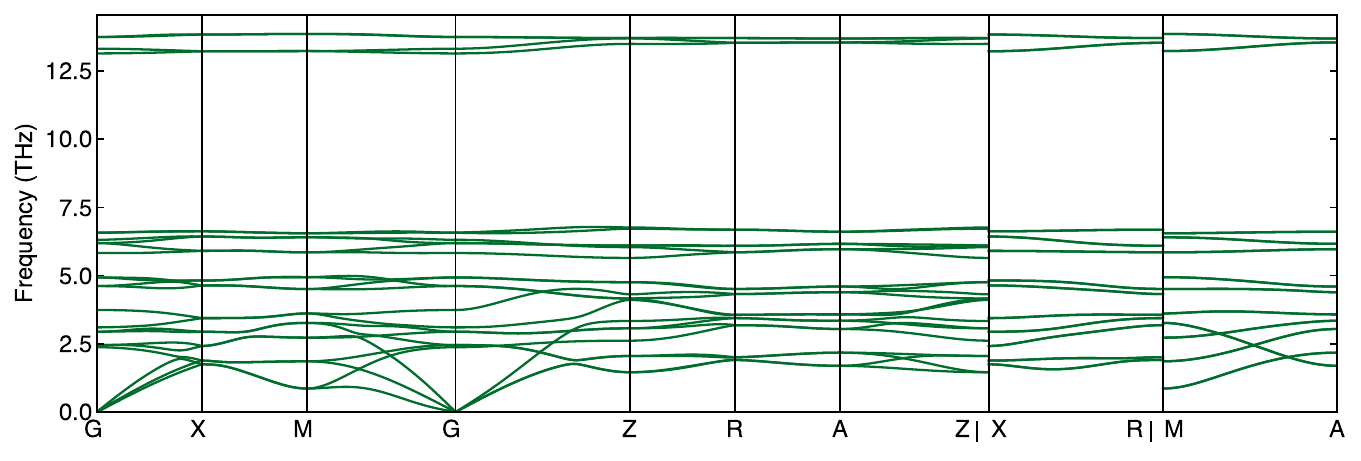}
    \caption{\ce{BaS3} ($P\Bar{4}2_1m$) phonon band structure}
    \label{fig:P3}
\end{figure}

\begin{figure}[H]
    \centering
    \includegraphics[width=\textwidth]{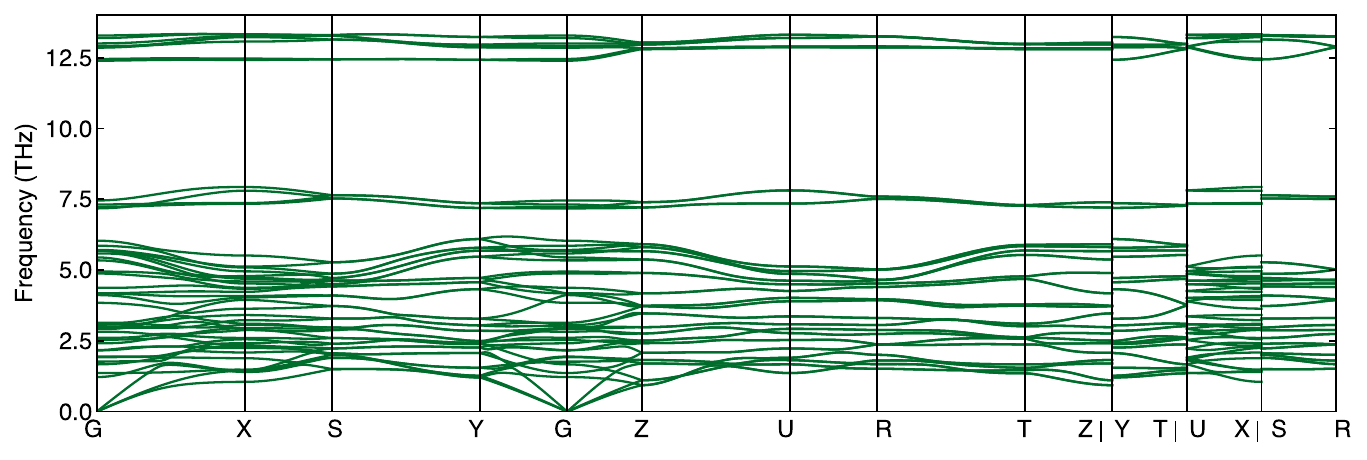}
    \caption{\ce{BaS3} ($P2_12_12$) phonon band structure}
    \label{fig:P4}
\end{figure}

\begin{figure}[H]
    \centering
    \includegraphics[width=\textwidth]{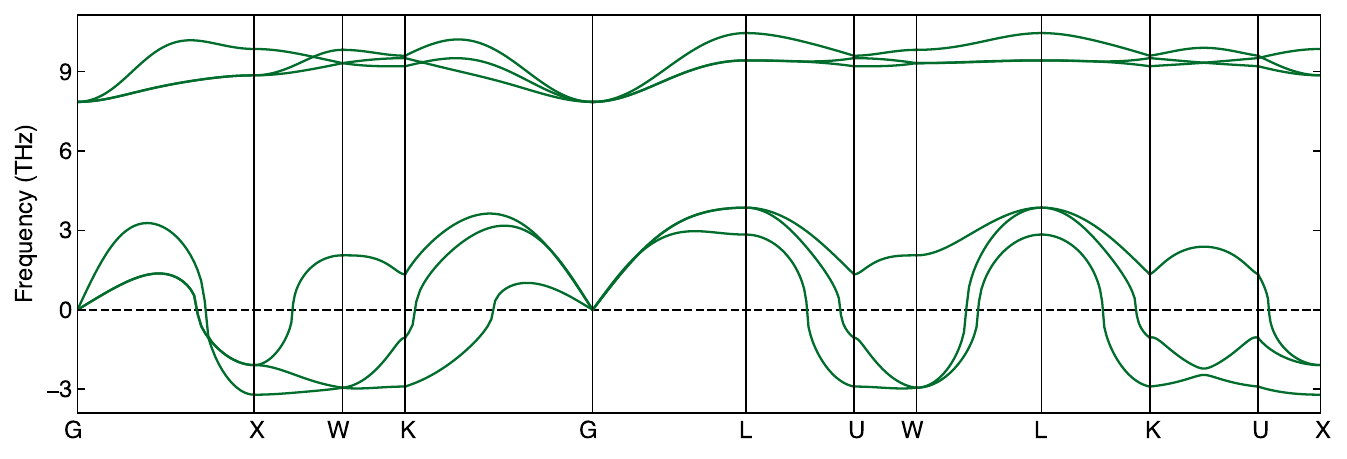}
    \caption{\ce{ZrS} ($Fm\Bar{3}$m) phonon band structure}
    \label{fig:P5}
\end{figure}

\begin{figure}[H]
    \centering
    \includegraphics[width=\textwidth]{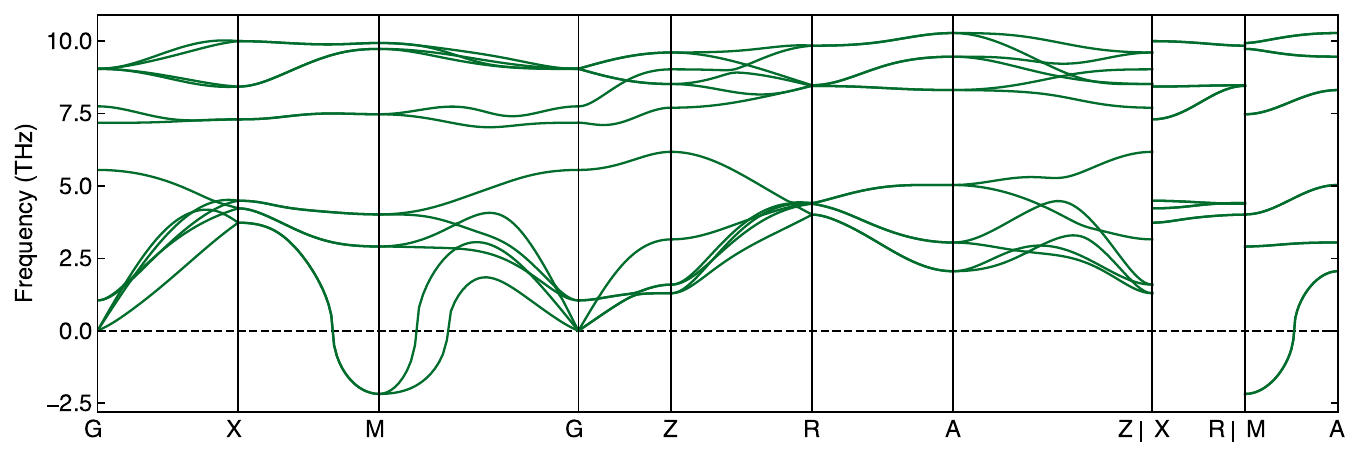}
    \caption{\ce{ZrS} ($P4/nmm$) phonon band structure}
    \label{fig:P6}
\end{figure}

\begin{figure}[H]
    \centering
    \includegraphics[width=\textwidth]{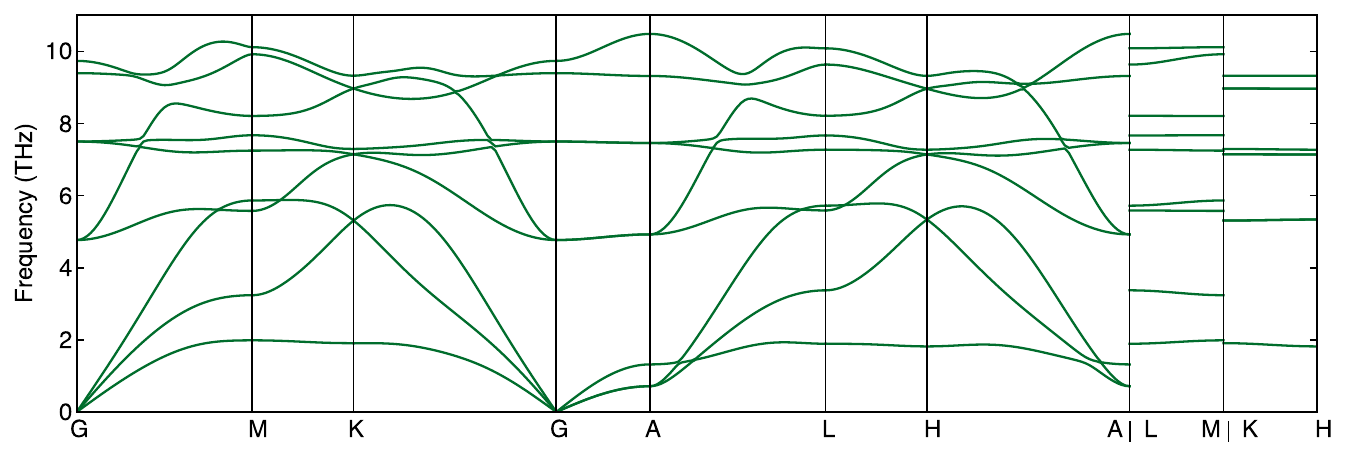}
    \caption{\ce{ZrS2} ($P\Bar{3}m1$) phonon band structure}
    \label{fig:P7}
\end{figure}

\begin{figure}[H]
    \centering
    \includegraphics[width=\textwidth]{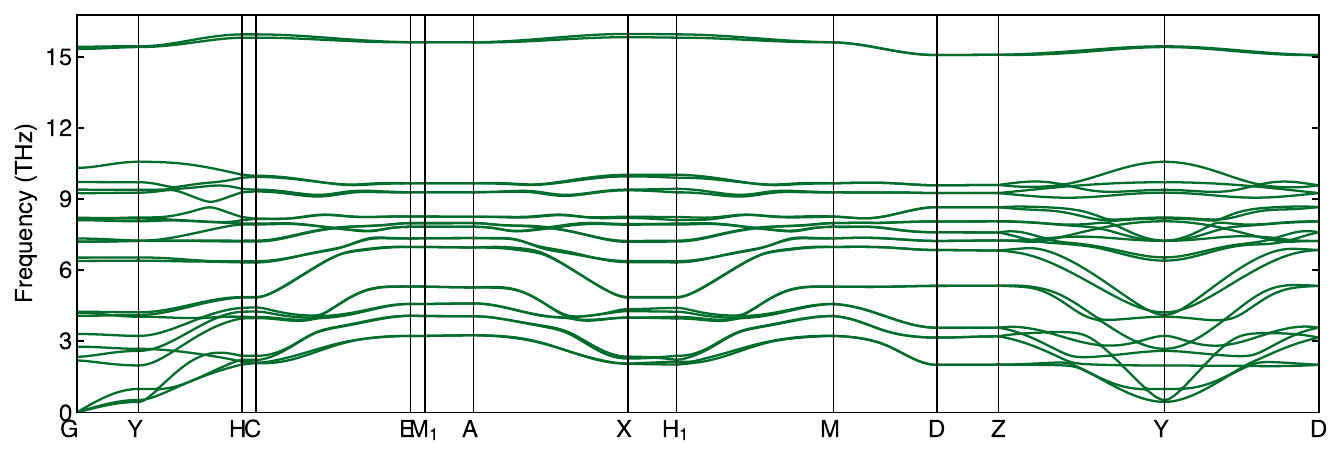}
    \caption{\ce{ZrS3} ($P2_1/m$) phonon band structure}
    \label{fig:P8}
\end{figure}

\begin{figure}[H]
    \centering
    \includegraphics[width=\textwidth]{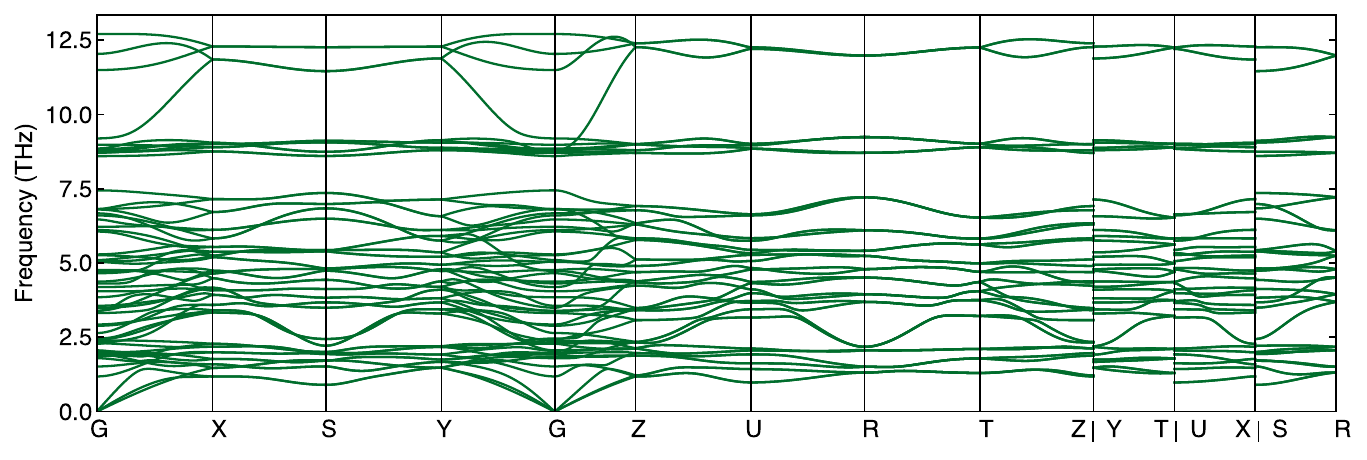}
    \caption{\ce{BaZrS3} ($Pnma$) phonon band structure}
    \label{fig:P9}
\end{figure}

Figures \ref{fig:P1} to \ref{fig:P9} present the phonon bandstructures for \ce{BaZrS3} and all of the binary materials selected for this study. The bandstructures are generated using first-principles lattice dynamics, as outlined in the methods section of the main text. They have been plotted using \texttt{sumo}.\cite{ganose2018sumo}
The only materials with a kinetic instability (ie, which have phonon mode(s) with an imaginary frequency) are the two \ce{ZrS} compounds. However we find that these are also thermodynamically unstable across the full temperature and pressure range considered in this study.


\newpage

\section{Formation of Ruddlesden-Popper phases}

As outlined in the main text, Ruddlesden-Popper (RP) phases are reported to form at high temperatures and using the same methodology as outlined in this paper we have demonstrated that they are energetically accessible above \SI{1000}{\kelvin}.\cite{kayastha2023high} An alternative RP formation mechanism is driven by a Zr or S deficit during synthesis. This will reduce the respective chemical potentials, allowing formation of the \ce{Zr}- and \ce{S}- deficit RP phases. By the same argument, the formation of RP-phases will be less energetically favourable when there is a sulfur-rich environment. Ruddlesden-Popper phases are not examined here as we focus our attention on perovskite formation at moderate temperatures and with an excess of sulfur.

\newpage

\section{\ce{Ba-Zr-S} ternary phase diagrams}

\begin{figure}[H]
    \centering
    \includegraphics[width=0.6\textwidth]{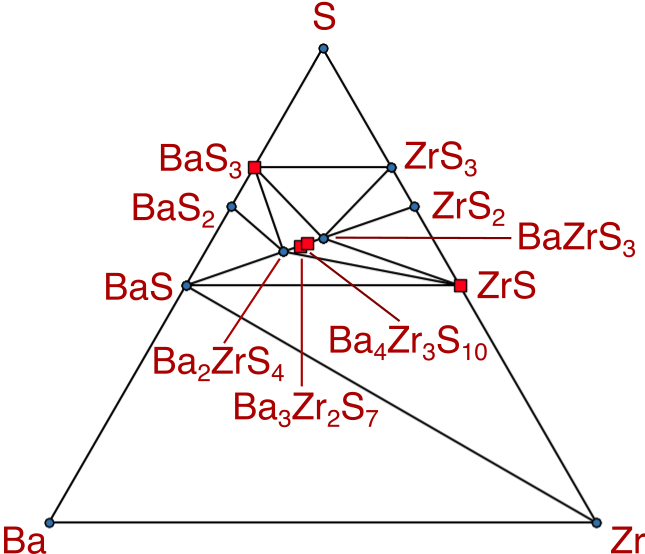}
    \caption{Phase diagram for the Ba-Zr-S system at 0K. The blue points indicate stable phases on the convex hull, whilst the red squares indicate an unstable phase lying within \SI{0.2}{\electronvolt}/atom of the convex hull. Unstable phases above this cutoff are not displayed.}
    \label{fig:BZS-PD-0}
\end{figure}

\begin{figure}[H]
    \centering
    \includegraphics[width=0.6\textwidth]{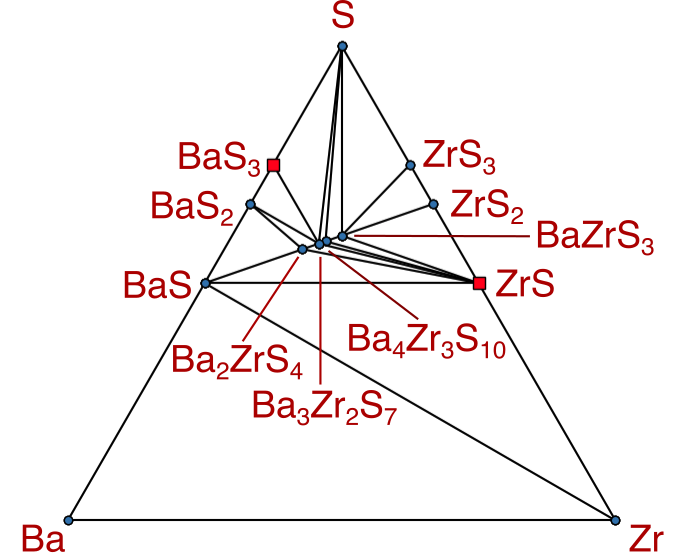}
    \caption{Phase diagram for the Ba-Zr-S system at 800K. The blue points indicate stable phases on the convex hull, whilst the red squares indicate an unstable phase lying within \SI{0.2}{\electronvolt}/atom of the convex hull. Unstable phases above this cutoff are not displayed.}
    \label{fig:BZS-PD-800}
\end{figure}

In Figures \ref{fig:BZS-PD-0} and \ref{fig:BZS-PD-800} we show the phase diagram for the Ba-Zr-S system at 0K and 800K respectively. The phase diagrams have been calculated using ThermoPot\cite{ThermoPot} and Pymatgen.\cite{Ong2008Li} Total energies (using the SCAN exchange-correlation functional) and vibrational contributions have been calculated as outlined in the Methods section of the main text.

\newpage

\newpage
\bibliography{si}